\documentclass[10pt,twocolumn,letterpaper]{article}

\usepackage[pagenumbers]{cvpr} 
\usepackage{graphicx}
\usepackage{amsmath}
\usepackage{amssymb}
\usepackage{booktabs}
\usepackage{acronym}
\usepackage{graphicx}
\usepackage{adjustbox}
\usepackage{multirow}
\usepackage{color, soul}
\usepackage{xcolor}
\usepackage{tikz}
\usepackage{enumitem}

\usepackage[pagebackref,breaklinks,colorlinks]{hyperref}

\usepackage[capitalize]{cleveref}
\crefname{section}{Sec.}{Secs.}
\Crefname{section}{Section}{Sections}
\Crefname{table}{Table}{Tables}
\crefname{table}{Tab.}{Tabs.}

\def\cvprPaperID{5703} 
\def\confName{CVPR}
\def\confYear{2022}

\newacro{lr}[LR]{low-resolution}
\newacro{hr}[HR]{high-resolution}
\newacro{sr}[SR]{super-resolution}
\newacro{cr}[CR]{compact-resolution}
\newacro{cnn}[CNN]{convolutional neural network}
\newacro{ar}[AR]{autoregressive}
\newacro{cgan}[cGAN]{conditional generative adversarial network}
\newacro{dct}[DCT]{discrete cosine transorm}
\newacro{oled}[OLED]{organic light-emitting diode}
\newacro{ace}[ACE]{adaptive contrast enhancement}
\newacro{race}[R-ACE]{residual-adaptive contrast enhancement}
\newacro{ssim}[SSIM]{structural similarity index measure}
\newacro{invean}[InvEAN]{invertible energy-aware network}
\newacro{inn}[INN]{invertible neural network}
\newacro{invdn}[InvDN]{invertible denoising network}
\newacro{jsdnss}[JSD-NSS]{Jensen Shannon divergence - natural scene statistics}
\newacro{lpips}[LPIPS]{learned perceptual image patch similarity}
\newacro{psnr}[PSNR]{peak signal-to-noise ratio}
\newacro{kld}[KLD]{KL divergence}
\newacro{vif}[VIF]{visual information fidelity}
\newacro{ghg}[GHG]{greenhouse gas}

\begin{document}
\setlength{\abovedisplayskip}{2pt}
\setlength{\belowdisplayskip}{2pt}
\title{
3R-INN: How to be climate friendly while consuming/delivering videos?}

\author{Zoubida Ameur\\
InterDigital R\&D, France\\
{\tt\small zoubida.ameur@interdigital.com}
\and
Claire-Hélène Demarty\\
InterDigital R\&D, France\\
{\tt\small claire-helene.demarty@interdigital.com}
\and
Daniel Ménard\\
INSA-Rennes, France\\
{\tt\small daniel.menard@insa-rennes.com}
\and
Olivier Le Meur\\
InterDigital R\&D, France\\
{\tt\small olivier.lemeur@interdigital.com}
}

\twocolumn[{%
\renewcommand\twocolumn[1][]{#1}%
\maketitle
\begin{center}
    \centering
    \captionsetup{type=figure}
    \includegraphics[width=\linewidth]{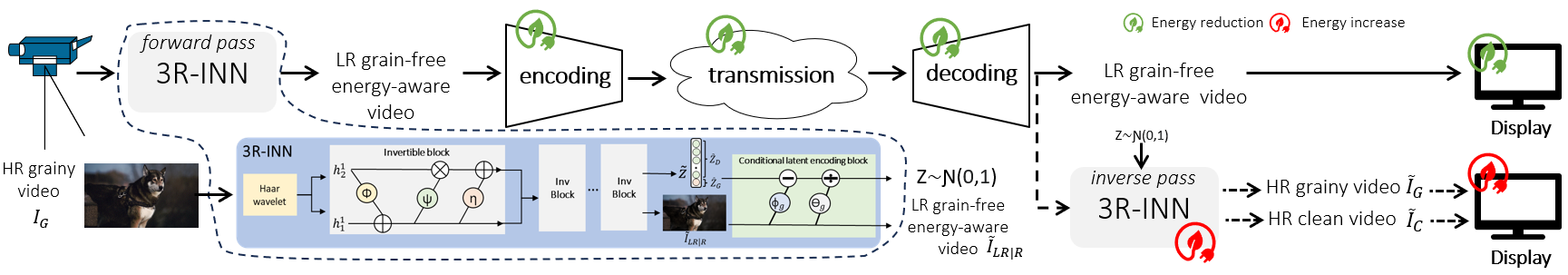}
    \captionof{figure}{3R-INN: End-to-end \textbf{energy-aware} video distribution chain by \textbf{R}emoving grain, \textbf{R}escaling and \textbf{R}educing display energy.}
    \label{fig:use_case}
\end{center}%
}]

\begin{abstract}
The consumption of a video requires a considerable amount of energy  during the various stages of its life-cycle. With a billion hours of video consumed daily, \textbf{this contributes significantly to the \ac{ghg} emission}. Therefore, reducing the end-to-end carbon footprint of the video chain, while preserving the quality of experience at the user side, is of high importance. To contribute in an impactful manner, we propose 3R-INN, a single light invertible network that does three tasks at once: given a \ac{hr} grainy image, it \textbf{R}escales it to a lower resolution, \textbf{R}emoves film grain and \textbf{R}educes its power consumption when displayed. Providing such a minimum viable quality content contributes to reducing the energy consumption during encoding, transmission, decoding and display. 3R-INN also offers the possibility to restore either the \ac{hr} grainy original image or a grain-free version, thanks to its invertibility and the disentanglement of the high frequency, and without transmitting auxiliary data. Experiments show that,while enabling significant energy savings for encoding (78\%), decoding (77\%) and rendering (5\% to 20\%), 3R-INN outperforms state-of-the-art film grain synthesis and energy-aware methods and achieves state-of-the-art performance on the rescaling task on different test-sets.

\end{abstract}
%
\section{Introduction}
Over 75\% of the world's global \ac{ghg} emissions comes from energy production, particularly from fossil fuels. The growing energy consumption of the media and entertainment (M\&E) industry, in particular streaming, strongly contributes to climate change, with more than 1.3\% of \ac{ghg} in 2020~\cite{carbontrust2021}. Therefore, M\&E industry has to move towards decarbonisation, energy efficiency and sustainability in all its stages, \eg, head-end (encoding), delivery (transmission) and end-user device (decoding and display). Taking apart the energy consumed while building the different necessary equipment, reduced energy consumption at the head-end translates into shorter encoding times and lower computing loads, while at the distribution level it translates into lower bit-rates. At the end-device level, significant gains can be achieved, as displays constitute the most power-hungry part of the whole chain~\cite{carbontrust2021}. In the specific case of emissive displays, \eg, \acp{oled}, the power consumption is directly pixel-wise and therefore dependent on the displayed content. Consequently, less energy-intensive images at display and shorter decoding times will also lead to lower energy consumption.  

The encoding and decoding times are related to the content resolution and complexity. Thus, downscaling the latter before encoding and upscaling it after decoding while preserving the same of quality of experience~\cite{9054716} is one straightforward solution to reduce the computational burden. Additionally, removing and modeling artistic noise, such as film grain, before encoding and synthesizing it after decoding, not only reduces encoding and decoding times, but also significantly reduces the bit-rate~\cite{norkin2018film}, while still preserving the artistic intent at the user side. Finally, as displays consume the largest proportion of the energy, providing energy-aware content, \ie, that will consume less when displayed, is of significant importance, at least for OLED displays. Several studies addressed this issue by investigating how to reduce the content brightness.

Because the climate change issue is pressing, we believe that having a global vision on how to reduce the overall energy consumption in the video chain is of the utmost importance. Therefore, in this paper, we propose an end-to-end energy reduction of the video distribution chain, while preserving a good quality of experience at the user side, by leveraging a deep learning \ac{inn}-based model, called 3R-INN. 

Prior to encoding a \ac{hr} image, our 3R-INN multi-task network \textbf{R}escales it to a lower resolution, \textbf{R}emoves film grain and \textbf{R}educes its power consumption when displayed, by some reduction rate $R$.  While saving energy along the video chain, 3R-INN also provides a visually-pleasant content intended to be displayed. In that sense, we follow the new paradigm proposed in~\cite{robinson2023less}, which promotes to target a minimum viable video quality for transported videos, but with the possibility to recover the original content, with the counter part that it will consume more.
The 3R-INN output corresponds to this viable video quality. Provided that it is accepted to run it in an inverse manner, thus consuming some energy, its invertible property allows to retrieve the \ac{hr} original version of the image. Furthermore, thanks to the modeling and disentanglement of the lost information in the forward pass, two versions, grainy and clean, of the original \ac{hr} image can be restored, 
without transmitting any auxiliary information. Because the energy consumed by applying any energy reduction processing should not be higher than the amount of saved energy, we also designed 3R-INN as a single light network, that could replace three separate and potentially heavier processings. 
In summary, our main contributions are five-folds:
\begin{itemize}[noitemsep,nolistsep,leftmargin=*]
\itemsep0em 
 \item a single light network for the three tasks of rescaling, removing grain and reducing the energy at display, dedicated towards saving energy in the whole video chain;
\item the provision of a visually pleasant, energy reduced version of the original image, and the capability to go back to the original \ac{hr} grainy and grain-free images with no transmission of additional metadata along the video chain;
\item a first end-to-end solution for reducing the energy consumption of the video chain;
\item the best method so far for synthesizing film grain with high fidelity, and with no need of auxiliary data;
\item the best method so far for building energy-aware images.
\end{itemize}

In the following, we first review the state-of-the-art for rescaling, film grain removal/synthesis and energy-aware images (Section~\ref{sec:related_work}), before detailing our proposed solution (Section~\ref{sec:proposed_approach}). In Section~\ref{sec:experiments}, we evaluate our method against state-of-the-art solutions. An ablation study is performed and we provide an energy-driven analysis of the use of 3R-INN on videos. In Section~\ref{sec:conclusion}, we draw conclusions and perspectives.
%
%
\section{Related work}
\label{sec:related_work}
\noindent \textbf{Rescaling }
%
The rescaling task helps saving resources, through the storage and transfer of downscaled versions of an original \ac{hr} image/video. Recovering the original resolution while having pleasant LR content can be very challenging. For these purposes, to maximize the restoration performance while producing visually pleasant \ac{lr} content, several works learn jointly the two tasks, \ie, downscaling and upscaling. In~\cite{kim2018task}, an auto-encoder-based framework learns the optimal \ac{lr} image that maximizes the reconstruction performance of the \ac{hr} image. In~\cite{sun2020learned}, a downscaling method with consideration on the upscaling process is proposed. The method is trained in an unsupervised manner, with no assumption on how the \ac{hr} image is downscaled, to learn the essential information for upscaling in an optimal way. Following a different paradigm, authors in~\cite{xiao2020invertible} model the down- and up-scaling processes using an invertible bijective transformation. In a forward pass, the framework performs the downscaling process by producing visually pleasing \ac{lr} images while capturing the distribution of the lost information using a latent variable that follows a specified distribution. Meanwhile, the upscaling process is made tractable such that the \ac{hr} image is reconstructed by inversely passing a randomly drawn latent variable with the \ac{lr} image through the network.


\noindent \textbf{Film grain removal and synthesis }
To better preserve film grain while compressing video content efficiently, it is classically removed and modeled before encoding and restored after decoding~\cite{gomila2003sei,norkin2018film}. 
Hence, dedicated methods for film grain removal are proposed, based on either temporal filtering~\cite{dai2010film}, spatio-temporal inter-color correlation filtering~\cite{hwang2013enhanced} or deep-learning encoder-decoder models~\cite{ameur2023deep}. On the other hand, several studies addressed the film grain synthesis task. In~\cite{newson2017stochastic}, a Boolean in-homogeneous model~\cite{stoyan2013stochastic} is used to model the grain, which corresponds to uniformly distributed disks.
In AV1 codec~\cite{norkin2018film}, film grain is modeled by an \ac{ar} method as well as by an intensity-based function to adjust its strength.
In VVC~\cite{radosavljevic2021implementation}, a method based on frequency filtering is used. The grain pattern is first modeled thanks to a \ac{dct} transform applied to the grain blocks corresponding to smooth regions, and further scaled to the appropriate level, by using a step-wise scaling function. In~\cite{ameur2023deep}, a \ac{cgan} that generates grain at different intensities is proposed. This model does not perform any analysis on the original grain for a reliable synthesis. In~\cite{ameur2023style}, a deep-learning framework is proposed which consists of a style encoder for film grain style analysis, a mapping network for film grain style generation, and a synthesis network that generates and blends a specific grain style to a given content in a content-adaptive manner.
%
%

\noindent \textbf{Energy-aware images }
Many works addressed the task of reducing the energy consumption of images while displayed on screens, especially for \ac{oled} displays. A first set of methods reduce the luminance through clipping or equalizing histograms~\cite{Kang_I2GEC_2009,Kang_IQPC_2015}. Other works directly scale the pixel luminance~\cite{Shin2019_ace, LeMeur2023deep, reinhard2023pixel}. The most promising methods leverage deep learning models, trained with a combination of loss functions that minimize the energy consumption while maintaining an acceptable perceptual quality. In~\cite{yin2020deep}, a deep learning model trained with a variational loss for simultaneously enhancing the visual quality and reducing the power consumption is proposed. Authors in~\cite{Shin2019_ace} describe a deep \ac{cnn} \ac{ace} network, that performs contrast enhancement of luminance scaled images. In~\cite{Nugroho2022r}, an improved version of ACE, called Residual-ACE (R-ACE), is proposed to infer an attenuation map instead of a reduced image. 
In~\cite{LeMeur2023deep}, authors revisit the R-ACE model to significantly reduce the complexity without compromising the performance. 
Different from the above methods,  an \ac{invean}~\cite{LeMeur2023invertible} produces invertible energy-aware images and allows to recover the original images if required.
\noindent \textbf{Invertible neural networks }
\acp{inn} learn the mapping $x=f(z)$, which is fully invertible as $z = f^{-1}(x)$, through a sequence of differentiable invertible mappings such as affine coupling layers~\cite{dinh2016density} and invertible 1×1 convolutional layers~\cite{kingma2018glow}. \acp{inn} have direct applications in ambiguous inverse problems by learning information-lossless mappings~\cite{zhao2021invertible,liu2021invertible,du2023hierarchical}. The lost information is captured by additional latent output variables. Thus, the inverse process is learned implicitly. A first application is the stenography, \ie, concealing images or a concatenation of multiple images~\cite{lu2021large, chen2023invertible}. In~\cite{zhao2021invertible}, an \ac{inn} is used to produce invertible grayscale images, where the lost color information is encoded into a set of Gaussian distributed latent variables. The original color version can be recovered by using a new set of randomly sampled Gaussian distributed variables as input, together with the synthetic grayscale, through the reverse mapping. Similarly, an \ac{invdn} transforms a noisy input into a \ac{lr} clean image and a latent representation containing noise in~\cite{liu2021invertible}. To discard noise and restore the clean image, \ac{invdn} replaces the noisy latent representation with another one sampled from a prior distribution during reversion. In~\cite{du2023hierarchical}, another \ac{inn}-based method further disentangles noise from the high frequency image information. 
%
%
\section{Proposed approach}\label{sec:proposed_approach}
With the target of reducing the overall energy consumption of the video transmission chain, our 3R-INN network performs three invertible tasks simultaneously: 1) film grain removal, 2) downscaling and 3) display energy reduction. This forward pass is run at the encoder side of the video chain as illustrated in Figure~\ref{fig:use_case}. 
From a \ac{hr} grainy image $I_G \in \mathbb{R}^{H \times W \times 3}$,  3R-INN outputs a visually pleasant grain-free \ac{lr} energy-aware image $\Tilde{I}_{LR|R} \in \mathbb{R}^{\frac{1}{2}H \times \frac{1}{2}W \times 3}$ with $R \in [0,1]$ being the energy reduction rate. To ensure the process invertibility and the bijective mapping, the lost information is captured in a latent variable $z$ distributed according to a standard Gaussian distribution $\mathcal{N}(0, 1)$. This can be formulated as: $[\Tilde{I}_{LR|R},z] = f_{\theta}(I_G)$  where $\Theta$ is the set of trainable parameters of the 3R-INN network $f$. $\Tilde{I}_{LR|R}$ is intended to be encoded, transmitted and displayed at the end-user device for an optimal energy consumption and quality of experience trade-off. 
During this process, the framework further disentangles the lost information into two parts, that come from film grain removal and the downscaling operation. This is done inside 3R-INN, by setting $\tilde{z}$ an internal representation of the lost information $z$ as $\tilde{z} = [\tilde{z}_D, \tilde{z}_G]$
with $\tilde{z}_D$ and $\tilde{z}_G$ representing losses due downscaling and film grain removal, respectively. 

In case the original content should be recovered, 3R-INN is run inversely at the decoder side (see Figure \ref{fig:use_case}), as follows: $\Tilde{I}_{G} = f_{\theta}^{-1}([\Tilde{I}_{LR|R},z])$. The \ac{hr} grainy version of the original content is then reconstructed with no need to transmit any auxiliary information in the video chain. Moreover, thanks to the film grain and high frequency loss disentanglement, $\tilde{z} = [\tilde{z}_D, \tilde{z}_G]$, the  framework is also able to generate a clean \ac{hr} version  $\Tilde{I}_{C}$ of the original content by setting $\tilde{z}_G = 0$. 
The overall architecture of the proposed framework is composed of three block types: one Haar Transformation block, several invertible blocks and a conditional latent encoding block, as illustrated in Figure \ref{fig:use_case}.
\subsection{Haar transform}
As removing film grain and downscaling an image significantly impacts high frequencies, it seems natural to first decompose the input \ac{hr} image into low and high-frequency components. For that purpose, we  chose the dyadic Haar wavelet transformation, similarly to~\cite{zhao2021invertible,xiao2020invertible}, because of its simplicity, efficiency and invertibility. Specifically, the Haar transform decomposes an input feature $f_{in} \in  \mathbb{R}^{H \times W \times C}$ into one low-frequency $f_{low} \in \mathbb{R}^{\frac{1}{2}H \times \frac{1}{2}W \times C}$ and three high-frequency $f_{high} \in \mathbb{R}^{\frac{1}{2}H \times \frac{1}{2}W \times 3C}$ sub-bands. $f_{low}$,  produced by an average pooling, represents the overall structure and coarse features of the image, while $f_{high}$ contains finer details in the vertical, horizontal and diagonal directions, corresponding to film grain and edges. 
This splitting strategy allows to separate very early in the process the low frequency components from the information we aim to suppress. 
$f_{low}$ and $f_{high}$ are then used as inputs of the following invertible blocks. 

\subsection{Invertible block}
As invertible blocks, we selected the coupling layer architecture proposed in~\cite{kingma2018glow}. A given input $h^{i}$ is composed of two parts $h^{i}_{1}$ and $h^{i}_{2}$, representing the three low-frequency and the nine high-frequency sub-bands of the color input channels $RGB$, respectively. These subbands are then processed by the $i^{th}$ invertible block as follows:
\begin{gather}
    h^{i+1}_{1} = h^{i}_{1} + \phi (h^{i}_{2})  \\ \nonumber
    h^{i+1}_{2} = h^{i}_{2} \odot \exp({\psi(h^{i+1}_{1})}) + \eta (h^{i+1}_{1})
\label{eq:1}
\end{gather}
where $\phi$, $\psi$ and $\eta$ are dense blocks~\cite{huang2017densely}. Given 
$[h^{i+1}_{1},h^{i+1}_{2}]$, the inverse transformation can be easily computed by:
\begin{gather}
    h^{i}_{2} =  (h^{i+1}_{2} - \eta (h^{i+1}_{1})) / \exp({\psi(h^{i+1}_{1})})\\ \nonumber
    h^{i}_{1} = h^{i+1}_{1} - \phi (h^{i}_{2})  
\label{eq:2}
\end{gather}
\subsection{Conditioned latent encoding block}
Invertible networks learn a bijective mapping between an input and an output distribution. In case of information loss, it is required to add a latent variable $\tilde{z}$ to ensure the invertible property. This latent variable is assumed to follow a standard Gaussian distribution which allows to avoid transmitting additional information for the reconstruction process, but also makes the reconstruction process case-agnostic. In our context, this would mean that the reconstruction of the \ac{hr} grainy ($\Tilde{I}_{G}$) or clean ($\Tilde{I}_{C}$) images would not rely on the a priori knowledge of the \ac{lr} image $\Tilde{I}_{LR|R}$. 
To overcome this limitation and to enable an image-adaptive reconstruction during the inverse pass, the lost information $\tilde{z}$ is transformed into a Gaussian distributed latent variable $z$ whose mean and variance are conditioned on $\Tilde{I}_{LR|R}$. This is done through the use of a latent encoding block inspired from~\cite{zhao2021invertible}, whose structure is a one-side affine coupling layer that normalizes $\tilde{z}$ into a standard Gaussian distributed variable $z$ as follows, with $\phi_g$ and $\theta_g$ being dense blocks: 
\begin{equation}
    z = (\tilde{z} - \phi_{g}(\Tilde{I}_{LR|R})) / \exp{(\theta_{g}(\Tilde{I}_{LR|R}))}
\end{equation}

The reverse mapping can be formulated as:
\begin{gather}
\small
    \tilde{z} = z \odot \exp{(\theta_{g} (\Tilde{I}_{LR|R}))} + \phi_{g}(\Tilde{I}_{LR|R})) 
\end{gather}
\subsection{Training objectives}
The training of 3R-INN is first performed for the film grain removal/synthesis and rescaling tasks only. The network is then fine-tuned by adding the energy reduction task. 

\subsubsection{Rescaling and film grain removal/synthesis tasks}
The \textbf{Forward Pass} optimization is driven by a fidelity loss $\mathcal{L}_{forw}$ to guarantee a visually pleasant clean \ac{lr} image $\Tilde{I}_{LR}$, and a regularization loss $\mathcal{L}_{reg}$ to guarantee that the latent variable $z$ follows a standard Gaussian distribution. 

To guide $f_\theta$ to generate $\Tilde{I}_{LR}$, a down-sampled image $I_{LR}$ of the \ac{hr} clean image $I_C$ is computed by a bicubic filter, and used as ground-truth to minimize $\mathcal{L}_{forw}$:
\begin{equation}
\small
    \mathcal{L}_{forw}(\Tilde{I}_{LR}, I_{LR}) = \frac{1}{N} \sum_{i=1}^{N} \vert \vert \Tilde{I}_{LR} - I_{LR}\vert \vert_{2}
\end{equation}

where N is the batch size. 
Second, the log-likelihood of the probability density function $p(z)$ of the standard Gaussian distribution is maximized as follows, with $D$ = dim($z$):
\begin{equation}
\small
    \mathcal{L}_{reg} = - \log (p(z)) = - \log (\frac{1}{(2 \pi)^{D/2}} \exp{(- \frac{1}{2} \vert \vert z \vert \vert^{2})} )
\end{equation}

The \textbf{Inverse Pass} optimization consists of two fidelity losses $\mathcal{L}_{back_G}$ and $\mathcal{L}_{back_C}$, to restore $\Tilde{I}_{G}$ and $\Tilde{I}_{C}$, respectively. For this purpose, the latent variable $z$ is first decoded into $\tilde{z}$ by the latent encoding block conditioned by the image $\Tilde{I}_{LR}$. Then the disentanglement of film grain ($G$) and fine details ($D$) is performed with $\tilde{z}= [\tilde{z}_{D},\tilde{z}_{G}]$.

$\Tilde{I}_{G}$ is reconstructed by considering all the information contained in $\tilde{z}$, \ie, related to film grain and fine details: 
\begin{equation}
\small
    \mathcal{L}_{back_G} (\Tilde{I}_{G}, I_G ) =  \frac{1}{N} \sum_{i=1}^{N} \vert \vert f^{- 1}(\Tilde{I}_{LR}, z])_{|[\tilde{z}_{D},\tilde{z}_{G}]} - I_G\vert \vert_{1}
\end{equation}

$\Tilde{I}_{C}$ is restored by considering only the subset $\tilde{z}_{D}$ of $\tilde{z}$, \ie, by using $\tilde{z}=[\tilde{z}_{D},\tilde{z}_{G}=0]$ as follows: 
\begin{equation}
\small
    \mathcal{L}_{back_C} (\Tilde{I}_{C}, I_C ) =  \frac{1}{N} \sum_{i=1}^{N} \vert \vert f^{ - 1}(\Tilde{I}_{LR}, z)_{|[\tilde{z}_{D},0]} -I_C \vert \vert_{1},
\end{equation}

For both fidelity losses, the $\ell_1$ norm is classically used as in~\cite{xiao2020invertible, liu2021invertible}. 
Finally, 3R-INN is trained for the first two tasks by minimizing the following weighted sum:
\begin{equation}
    \mathcal{L}_{total}= \lambda_{1}  \mathcal{L}_{forw} + \lambda_{2}  \mathcal{L}_{reg} + \lambda_{3}  \mathcal{L}_{back_C} + \lambda_{4}  \mathcal{L}_{back_G}  
    \label{eq:total1}
\end{equation}

\subsubsection{Energy-aware task}
Learning the energy-aware task needs to already have the model converged regarding the removal of grain and the downscaling. Thus, instead of directly learning all tasks altogether, we fine-tune 3R-INN during the forward pass with additional power and fidelity losses, $\mathcal{L}_{pow}$  and $\mathcal{L}_{SSIM}$, to output an energy-aware grain-free \ac{lr} image $\Tilde{I}_{LR|R}$, \ie, its power consumption is reduced by $R$ compared to the power consumption of $I_{LR}$. 
Contrary to most works computing energy aware images, that assume a linear relationship between the power consumption $P_Y$ of an image and its linearized luminance~\cite{Nugroho2022r}, we follow the model from~\cite{Demarty2023} dedicated to RGBW OLED screens, and compute $P_{RGBW}$ as the sum of the power consumed by the four individual R, G, B, W leds.  
Similarly to~\cite{LeMeur2023invertible}, the following power loss is then minimized:
\begin{equation}
    \mathcal{L}_{pow} = \vert \vert \tilde{P}_{RGBW} - (1 - R) \times P_{RGBW} \vert \vert_{1}
\end{equation}
where $(1 - R) \times P_{RGBW}$ is the desired target power and $\Tilde{P}_{RGBW}$ the power of $\Tilde{I}_{LR|R}$.

To ensure a better visual quality of the energy-aware images, a \ac{ssim} loss is added and minimized as follows:
\begin{equation}
    \mathcal{L}_{SSIM} =1 - SSIM(\Tilde{I}_{LR|R}, I_{LR})
\end{equation}

As the inverse pass objectives remains exactly the same, the total loss minimized in the fine-tuning stage is:
\begin{gather}
    \mathcal{L}_{finetuned}= \mathcal{L}_{total}  + \lambda_{5}  \mathcal{L}_{pow} + \lambda_{6} \mathcal{L}_{SSIM}
    \label{eq:total2}
\end{gather}

\section{Experiments}
\label{sec:experiments}
%
%
\subsection{Training details}
During training, we use the DIV2K training set~\cite{agustsson2017ntire} from the FilmGrainStyle740K dataset~\cite{ameur2023style}, which contains pairs of corresponding images with and without grain. To complement the DIV2K validation set, we evaluate 3R-INN on the BSDS300 test set~\cite{martin2001database} and Kodak24 dataset~\cite{franzen1999kodak}, which were augmented to add grainy versions of the images, by following the same process as in the FilmGrainStyle740K dataset\footnote{The dataset will be made publicly available upon acceptance.}.
Input images were randomly cropped into $144 \times 144$ and augmented by applying random horizontal and vertical flips. Other training parameters are: Adam optimizer \cite{Adam1,Adam2} with $\beta_{1}$ = 0.9, $\beta_{2}$ = 0.999; mini-batch size of 16; 500k (training of the first two tasks) + 5k (energy-aware fine-tuning) iterations; learning rate  initialized as $2\text{e-}4$ and halved at [100k, 200k, 300k, 400k] mini-batch updates. Hyper-parameters 
are set to: $(\lambda_{1}, \lambda_{2}, \lambda_{3}, \lambda_{4}, \lambda_{5}, \lambda_{6}) =(40, 1, 1, 1, 1e10,1e4)$ and eight successive invertible blocks are used. 
Scale and shift coefficients are learned through a five-layer densely connected convolutional block. Each convolutional filter is of size $3\times 3$, with padding 1, followed by a leaky ReLU activation layer with negative slope set to 0.2. The intermediate channel number of the convolutional blocks is fixed to 32. Dimensions of $\tilde{z}_D$ and $\tilde{z}_G$ were set to (8, 1), respectively.
\begin{table}[h!]
\caption{Comparison between generated \ac{lr} clean images $\Tilde{I}_{LR|R=0}$ and a bicubic rescaling of the \ac{hr} clean image as ground-truth.}
\begin{adjustbox}{max width=\linewidth}
\begin{tabular}{l|cc|cc|cc}
\toprule
\multirow{2}{*}{Method} & \multicolumn{2}{c}{DIV2K} & \multicolumn{2}{c}{BSDS300} & \multicolumn{2}{c}{Kodak24} \\
                        & PSNR $\uparrow$ & SSIM $\uparrow$ & PSNR $\uparrow$ & SSIM $\uparrow$ & PSNR $\uparrow$ & SSIM $\uparrow$\\\midrule
IRN~\cite{xiao2020invertible} &39.06 &0.942  &38.95 &0.953  &38.75 &0.947 \\
Ours &\textbf{39.63} &\textbf{0.951} &\textbf{39.79} &\textbf{0.964} &\textbf{39.71} &\textbf{0.957}

\\ \bottomrule          
\end{tabular}
\end{adjustbox}
\label{tab:luma_lr_all}
\end{table}

\begin{figure}[h!]
    \centering
    \begin{subfigure}[]{0.32\linewidth}
        \includegraphics[width=\linewidth]{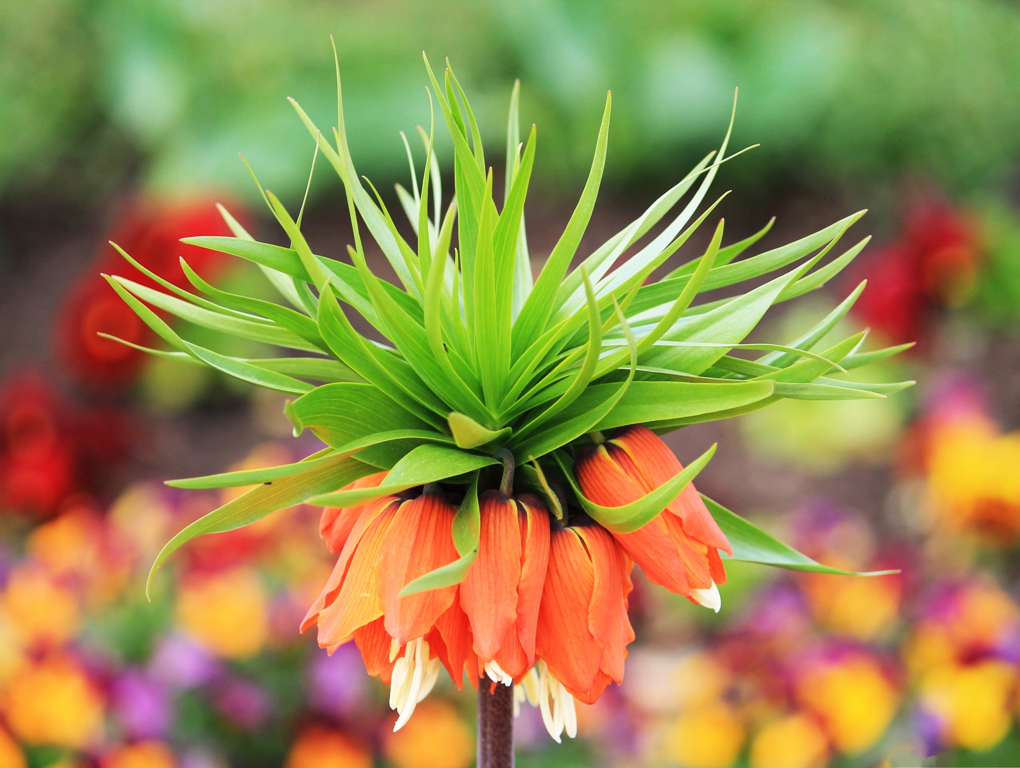}
        \caption*{Ground-truth\\ (bicubic)}
    \end{subfigure}
    \begin{subfigure}[]{0.32\linewidth}
        \includegraphics[width=\linewidth]{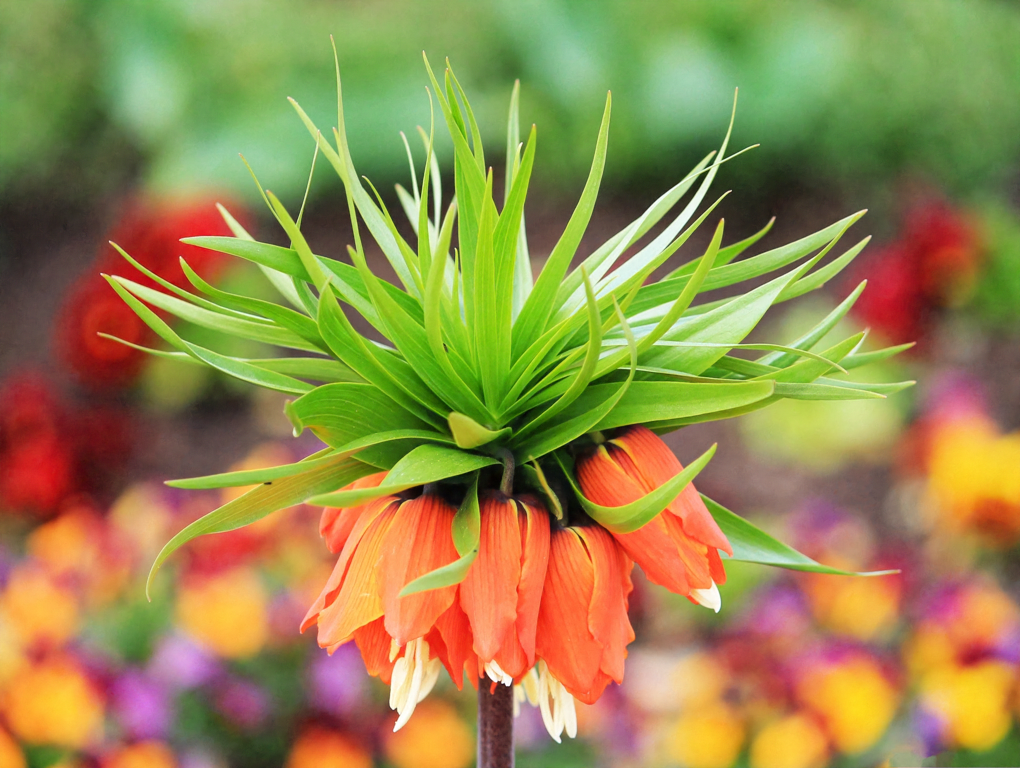}
        \caption*{IRN \\ (PSNR = 43.10 dB)}
    \end{subfigure}
    \begin{subfigure}[]{0.32\linewidth}
        \includegraphics[width=\linewidth]{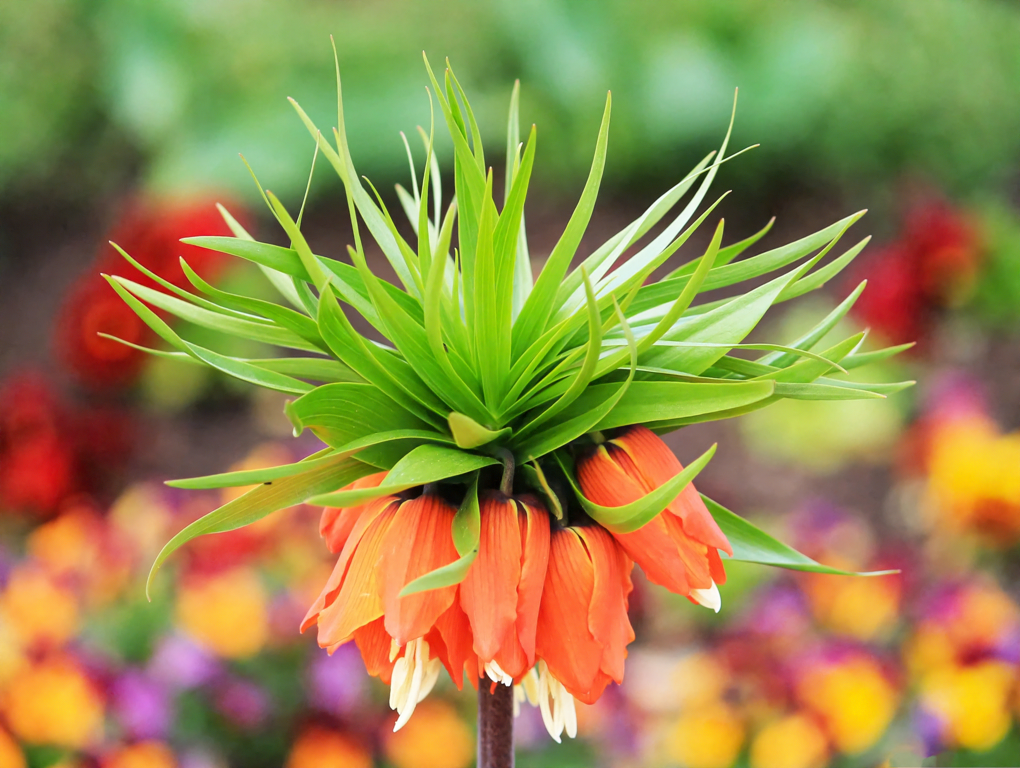}
        \caption*{Ours \\ (PSNR = 43.14 dB)}
    \end{subfigure}
\caption{Comparison between a bicubic downscaling, IRN and the generated clean \ac{lr} image $\Tilde{I}_{LR|R=0}$.}
\label{fig:removal_lr}
\end{figure}
\begin{table*}[t!]
\centering
\caption{PSNR-Y and SSIM quality scores for the energy-aware task for four energy reduction rate $R$. Results of the proposed method are presented for two power consumption models, \ie $P_{Y}$ (to be comparable to state-of-the-art methods) and $P_{RGBW}$, corresponding to RGB and RGBW OLED screens, respectively.}
\begin{adjustbox}{max width=\textwidth}
\begin{tabular}{l|c|cccc|cccc|cccc}
\toprule
\multirow{2}{*}{Method} &\multirow{2}{*}{Nb parameters} & \multicolumn{4}{c}{DIV2K}  & \multicolumn{4}{c}{BSDS}  & \multicolumn{4}{c}{Kodak24}                                                                                     \\ \cline{3-14}
   &    &R=5\%   &R=20\%   &R=40\%   &R=60\% &R=5\%   &R=20\%   &R=40\%   &R=60\% &R=5\%   &R=20\%   &R=40\%   &R=60\%\\ \midrule
LS &- &39.34/\textbf{0.999}    &27.01/0.991   &20.33/0.958   &16.06/0.877   &39.64/\textbf{0.999}   &27.31/0.990   &20.67/0.955   &16.35/0.867 &39.38/\textbf{0.999} &27.05/0.991   &20.41/0.957   &16.09/0.875  \\
R-ACE~\cite{Nugroho2022r} & 41K &41.53/0.995    &26.59/0.967   &20.05/0.901   &15.92/0.788   &40.55/0.997   &26.90/0.978   &20.24/0.915   &16.12/0.806 &40.70/0.997  &26.74/0.983   &20.08/0.930   &15.98/0.830   \\
DeepPVR~\cite{LeMeur2023deep} &4K &39.37/0.996    &27.12/0.983   &21.04/0.952   &15.81/0.890   &39.63/0.997   &27.53/0.989   &21.13/0.959   &16.36/0.894 &39.27/0.997   &27.17/0.989   &20.61/0.955   &16.00/0.892  \\

InvEAN~\cite{LeMeur2023invertible} &806K &-    &27.75/\textbf{0.994}   &21.17/0.973   &17.07/0.932   &-   &28.25/0.993   &21.74/0.973   &17.72/0.931 &- &27.92/0.993 &21.42/0.973   &17.37/0.932\\ \midrule

Ours ($P_{Y}$)  &1.7M & 39.55/0.987 & 27.32/0.980 & 20.62/0.949 & 16.43/0.883 & 40.06/0.994 & 27.65/0.986 & 20.94/0.955 & 16.77/0.883 & 40.02/0.992 & 27.43/0.985
& 20.70/0.954 & 16.51/0.886 \\
Ours ($P_{RGBW}$) &1.7M &\textbf{47.68}/0.998 &\textbf{38.02}/0.993 &\textbf{29.15/0.974} &\textbf{23.66/0.945} &\textbf{48.33/0.999} &\textbf{38.36/0.995} &\textbf{30.47/0.983} &\textbf{24.96/0.961} &\textbf{47.47}/0.998 &\textbf{37.39/0.994}
&\textbf{29.63/0.982} &\textbf{24.18/0.958}

\\\bottomrule

\end{tabular}
\label{tab:luma_energy_comparison}
\end{adjustbox}
\end{table*}

\begin{figure*}[t!]
    \centering
    \begin{subfigure}[]{0.15\linewidth}
        \includegraphics[width=\linewidth]{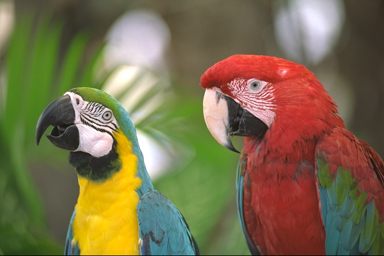}
    \end{subfigure}
    \begin{subfigure}[]{0.15\linewidth}
        \includegraphics[width=\linewidth]{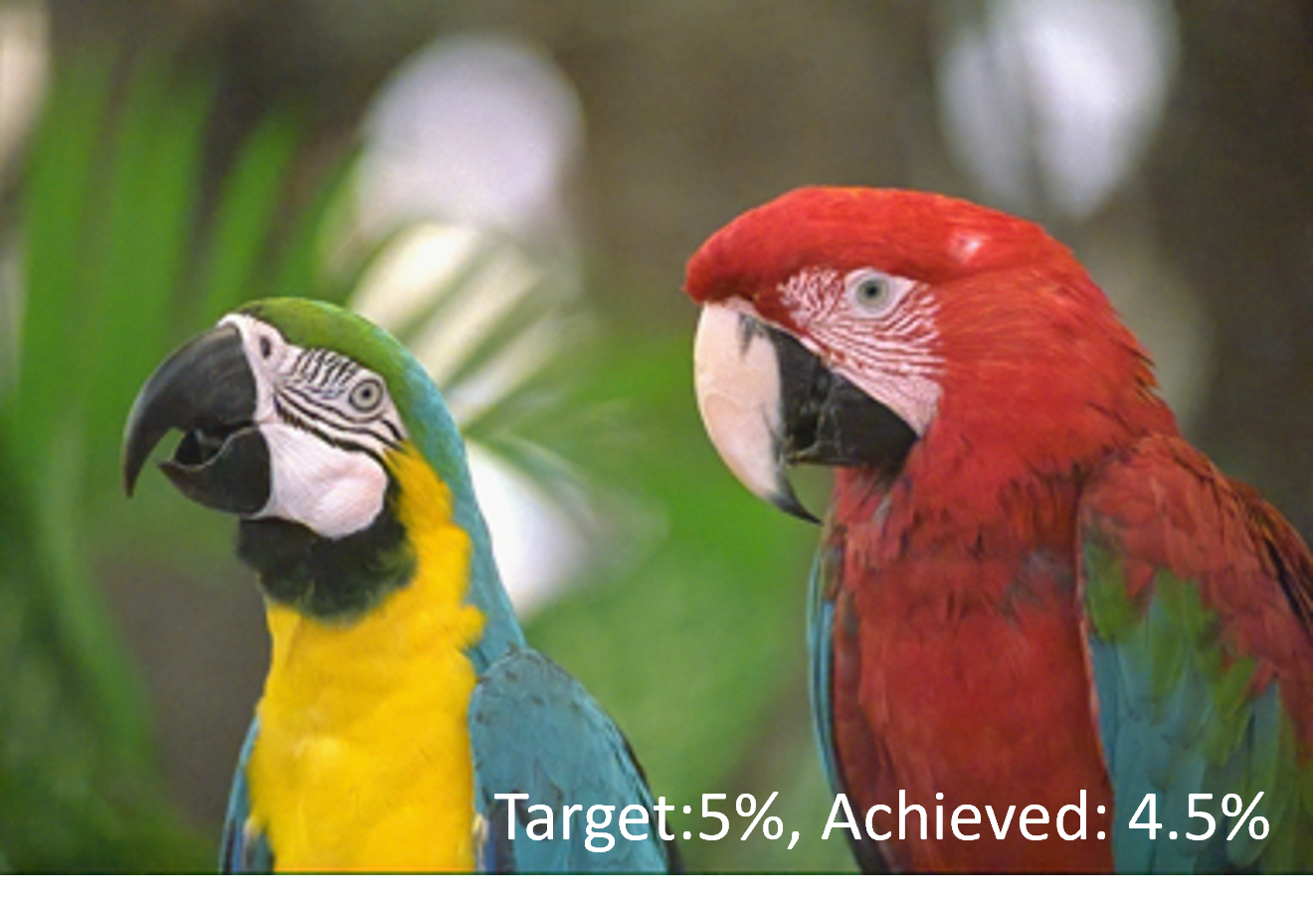}
    \end{subfigure}
    \begin{subfigure}[]{0.15\linewidth}
        \includegraphics[width=\linewidth]{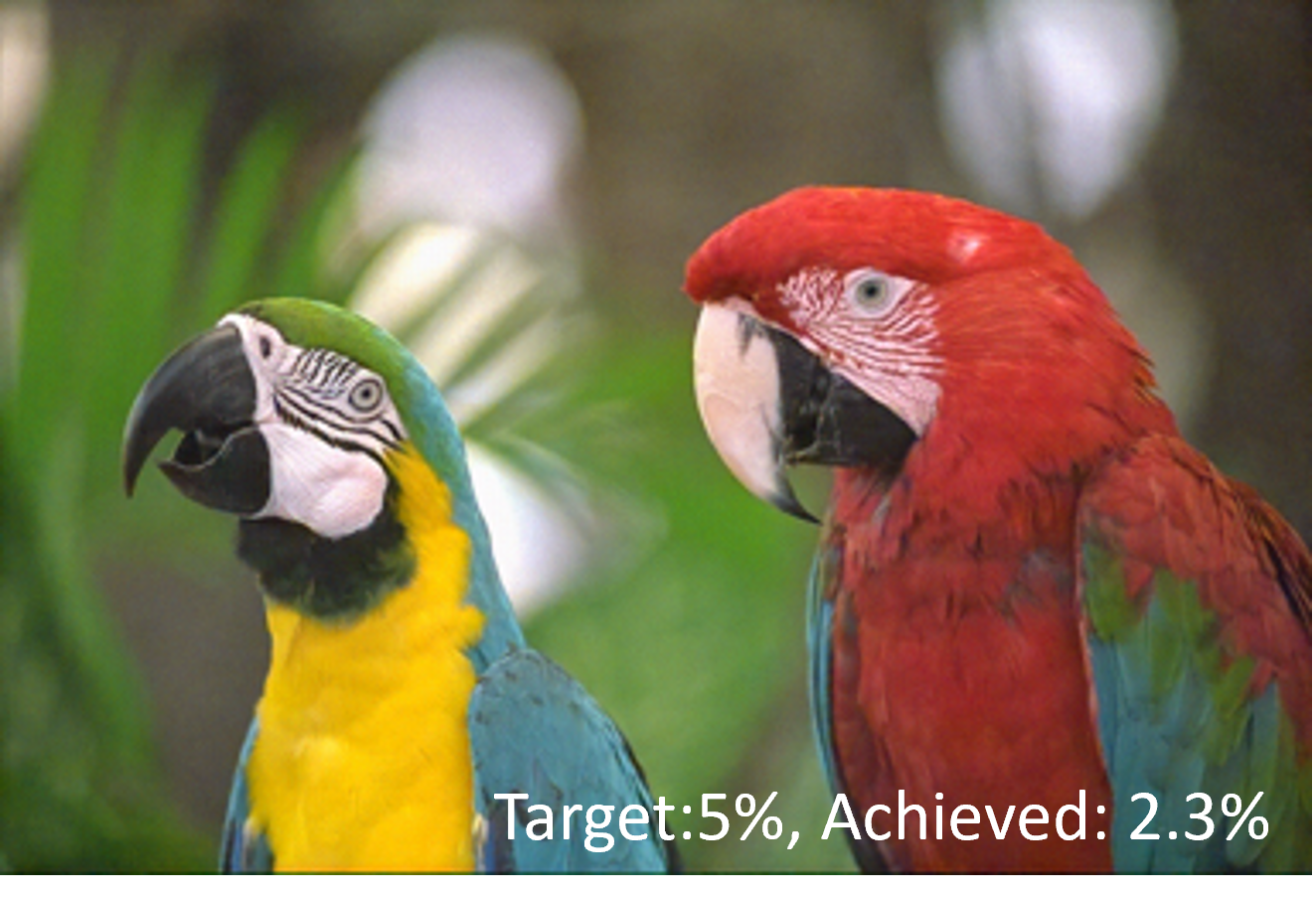}
    \end{subfigure}
    \begin{subfigure}[]{0.15\linewidth}
        \includegraphics[width=\linewidth]{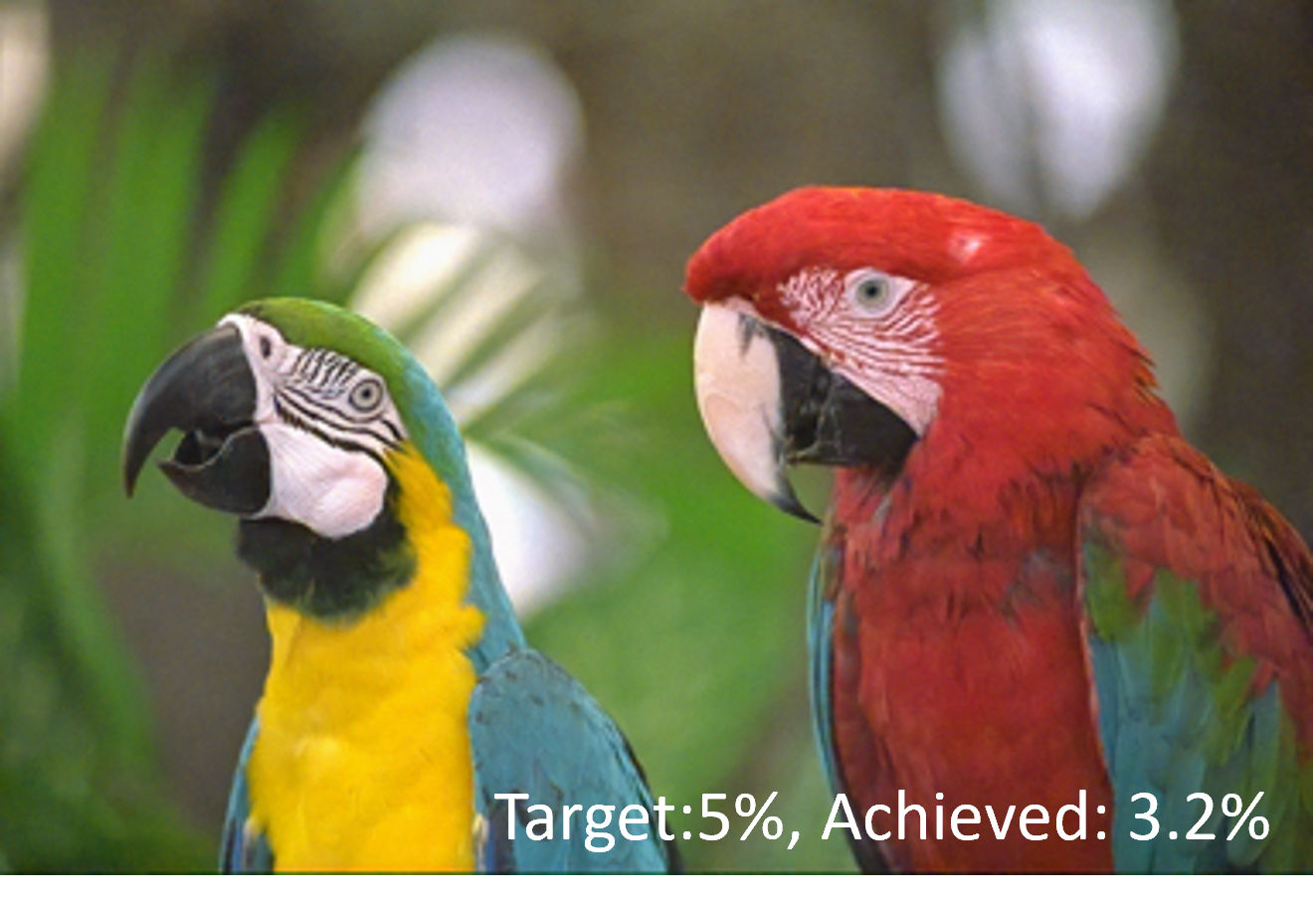}
    \end{subfigure}
    \begin{subfigure}[]{0.15\linewidth}
        \includegraphics[width=\linewidth]{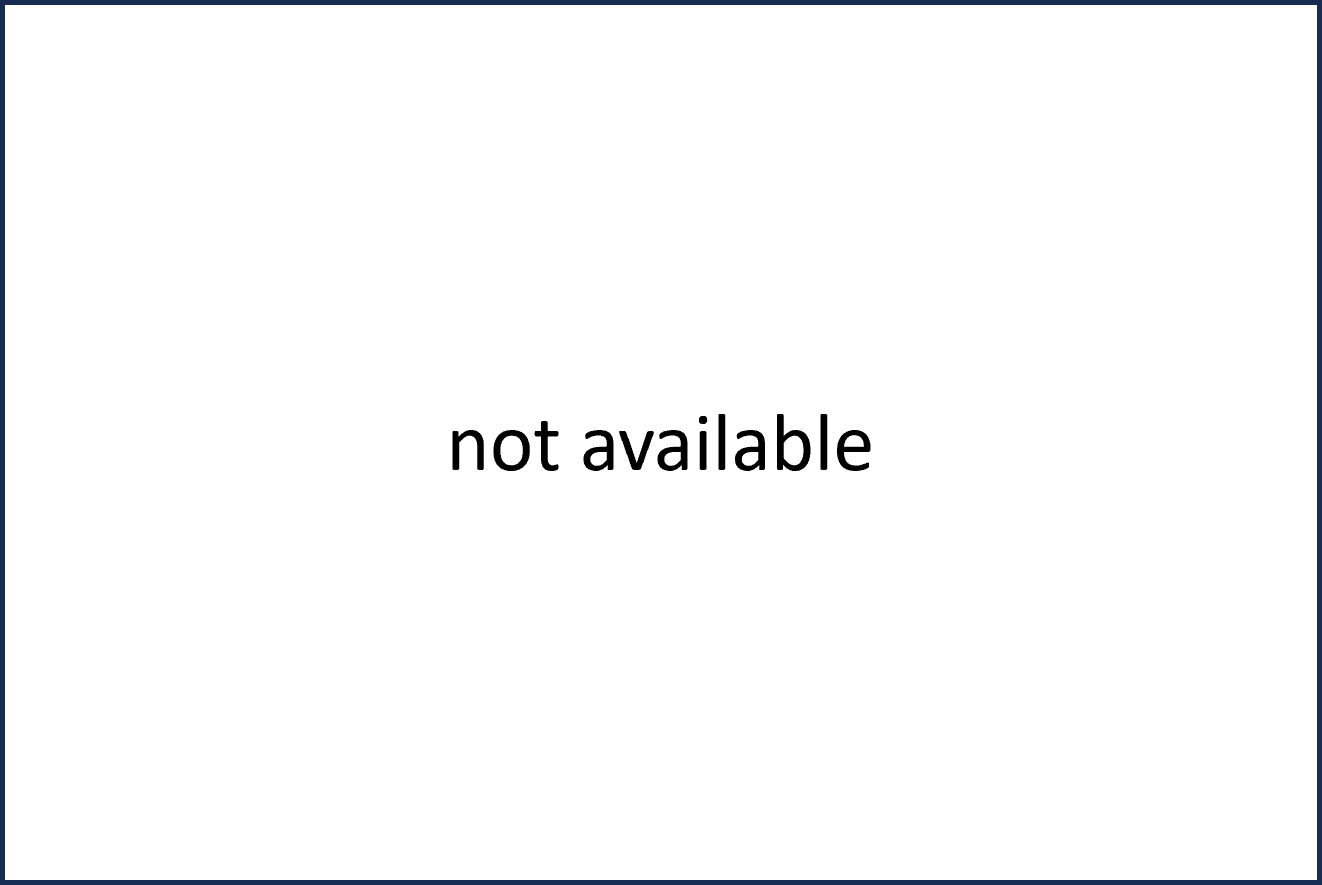}
    \end{subfigure}
    \begin{subfigure}[]{0.15\linewidth}
        \includegraphics[width=\linewidth]{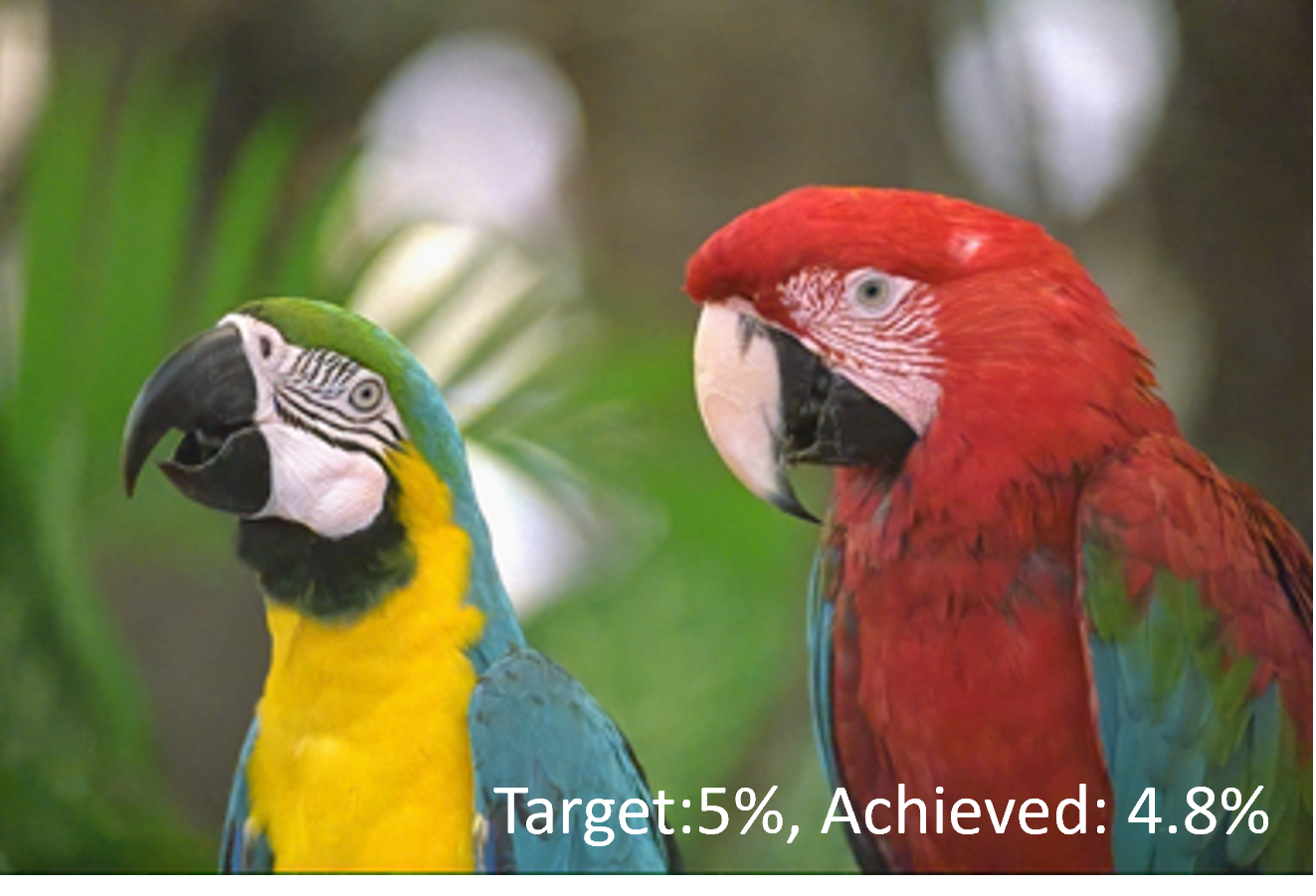}
    \end{subfigure}

    \begin{subfigure}[]{0.15\linewidth}
        \includegraphics[width=\linewidth]{figures/energy_lr/real/kodim23_combinaison_211_LR_ref.png}
    \end{subfigure}
    \begin{subfigure}[]{0.15\linewidth}
        \includegraphics[width=\linewidth]{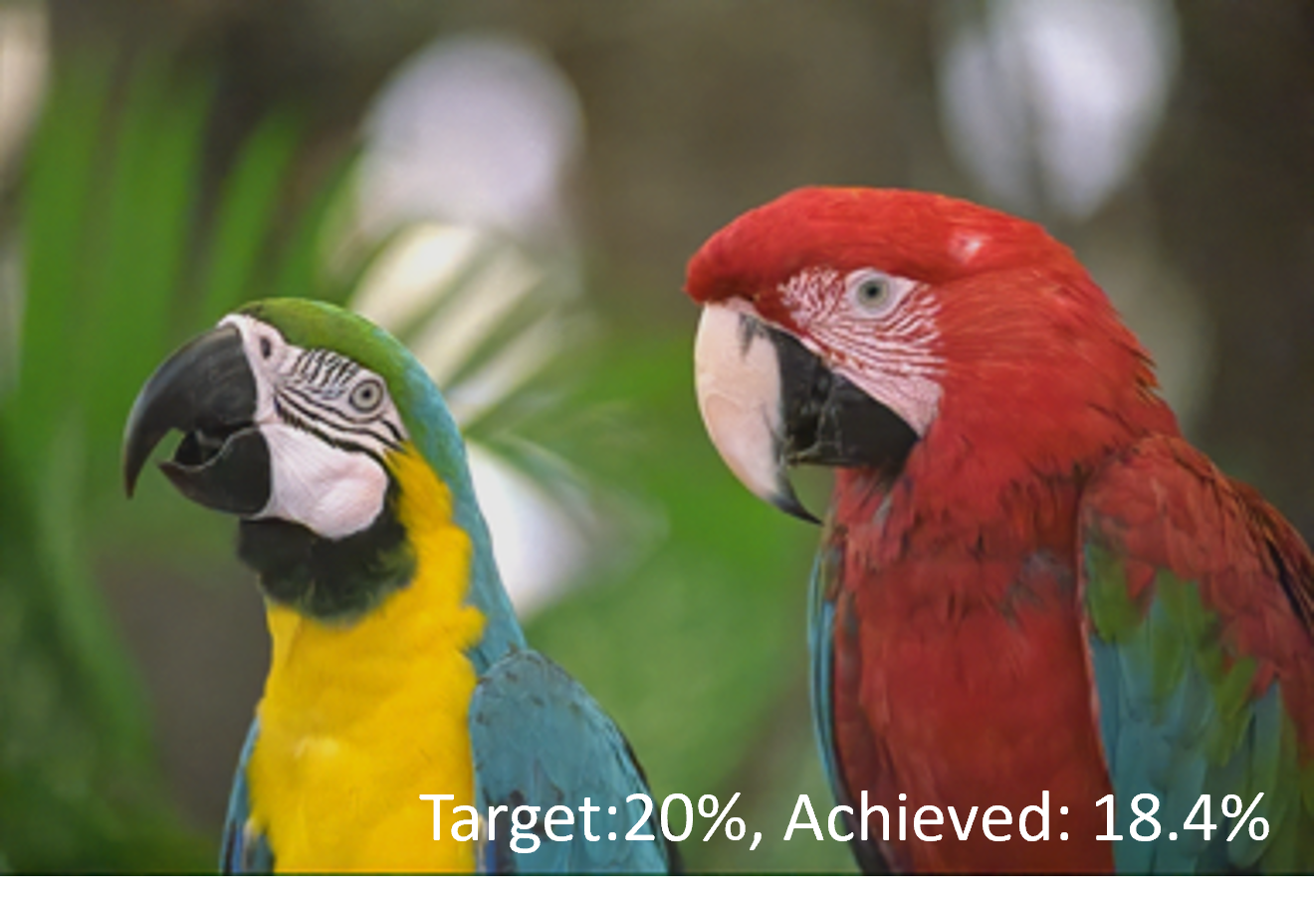}
    \end{subfigure}
    \begin{subfigure}[]{0.15\linewidth}
        \includegraphics[width=\linewidth]{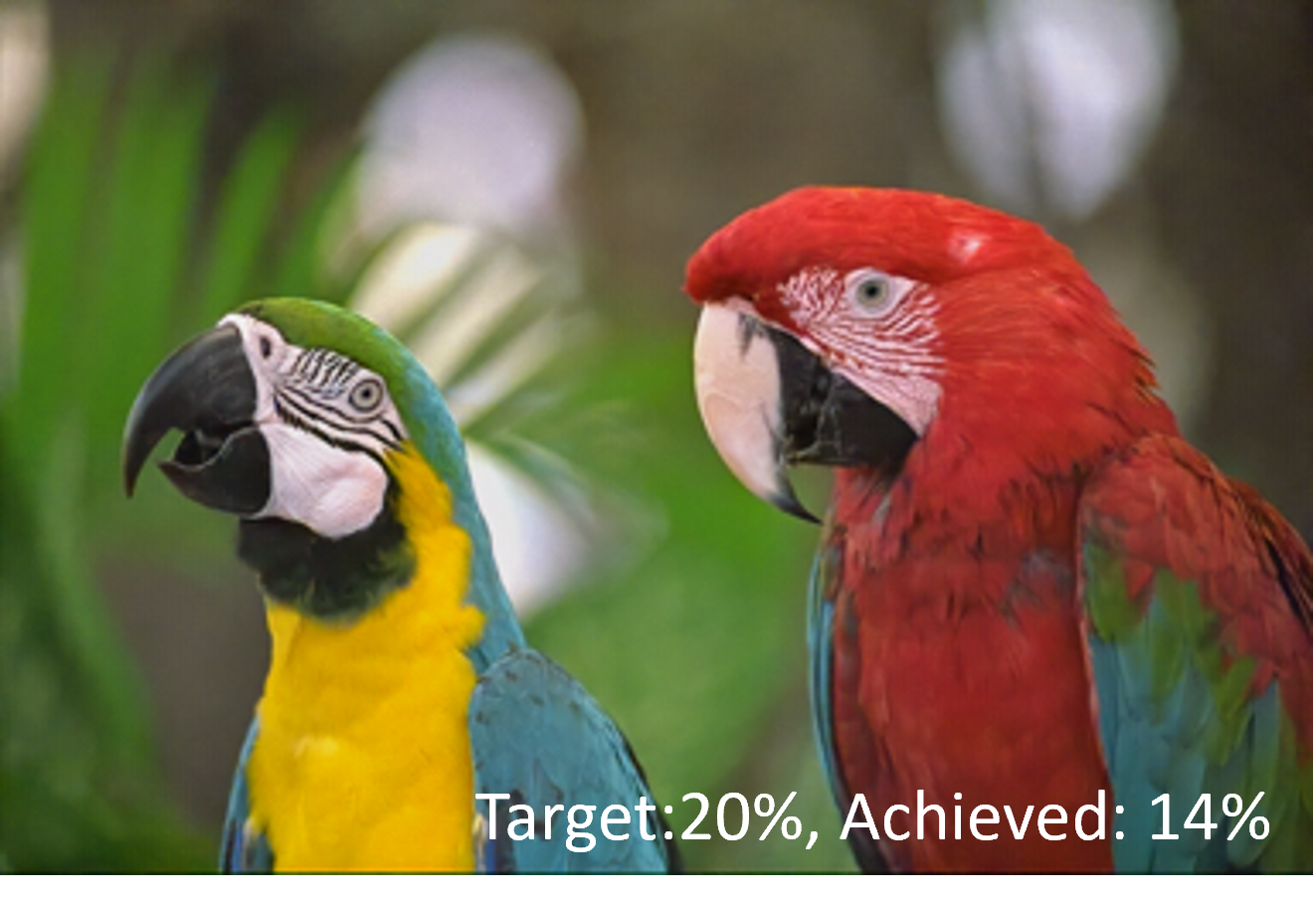}
    \end{subfigure}
    \begin{subfigure}[]{0.15\linewidth}
        \includegraphics[width=\linewidth]{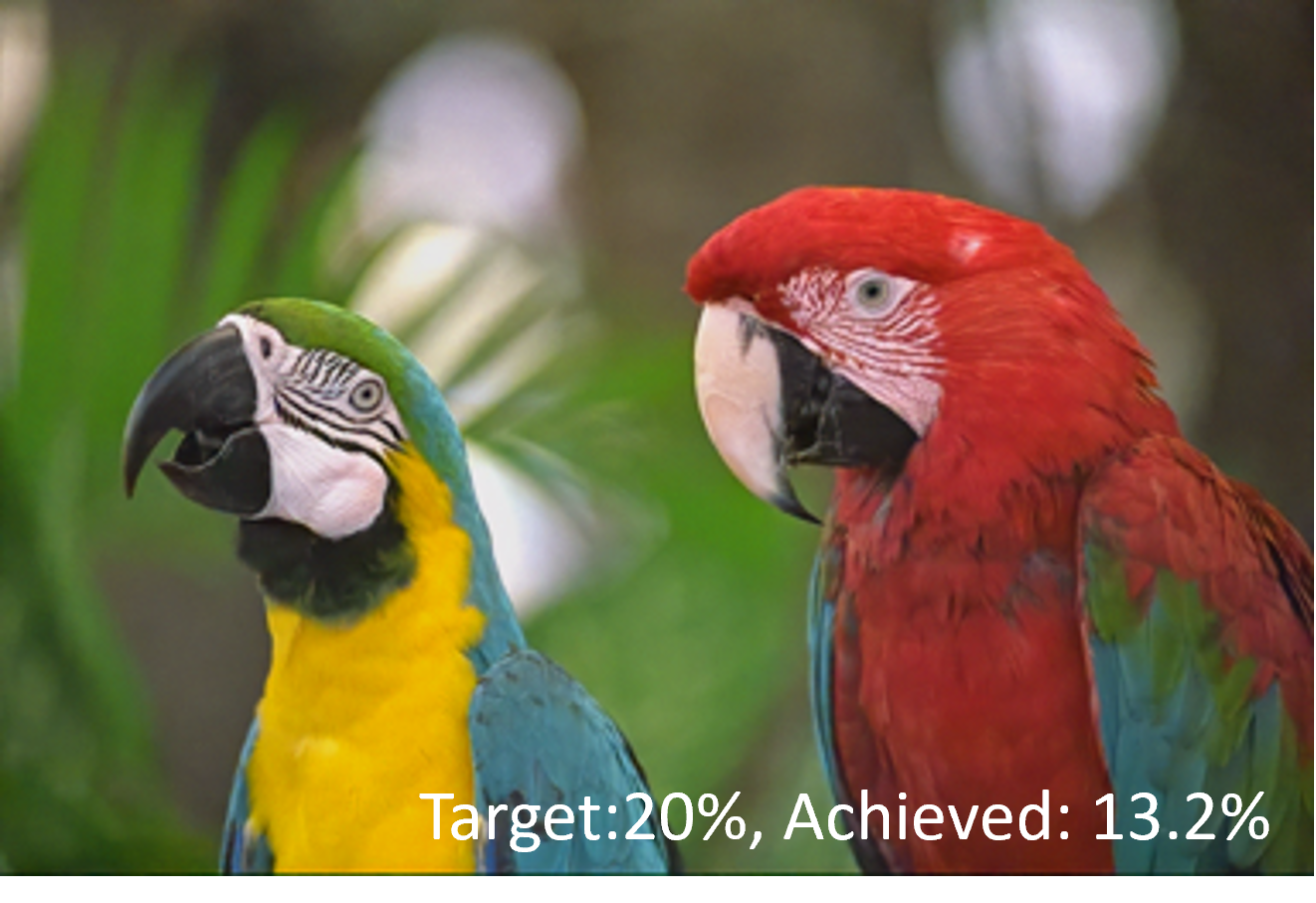}
    \end{subfigure}
    \begin{subfigure}[]{0.15\linewidth}
        \includegraphics[width=\linewidth]{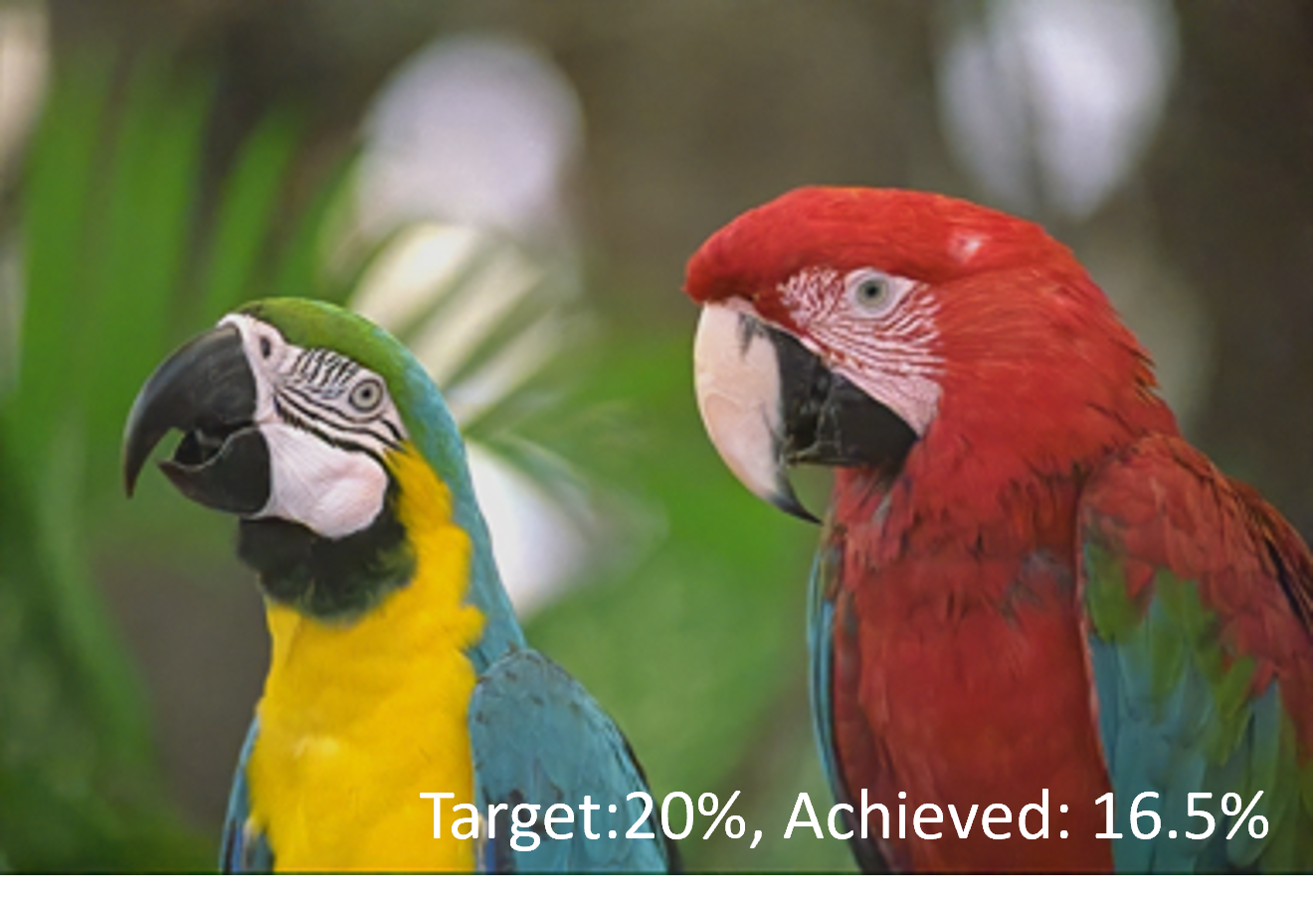}
    \end{subfigure}
    \begin{subfigure}[]{0.15\linewidth}
        \includegraphics[width=\linewidth]{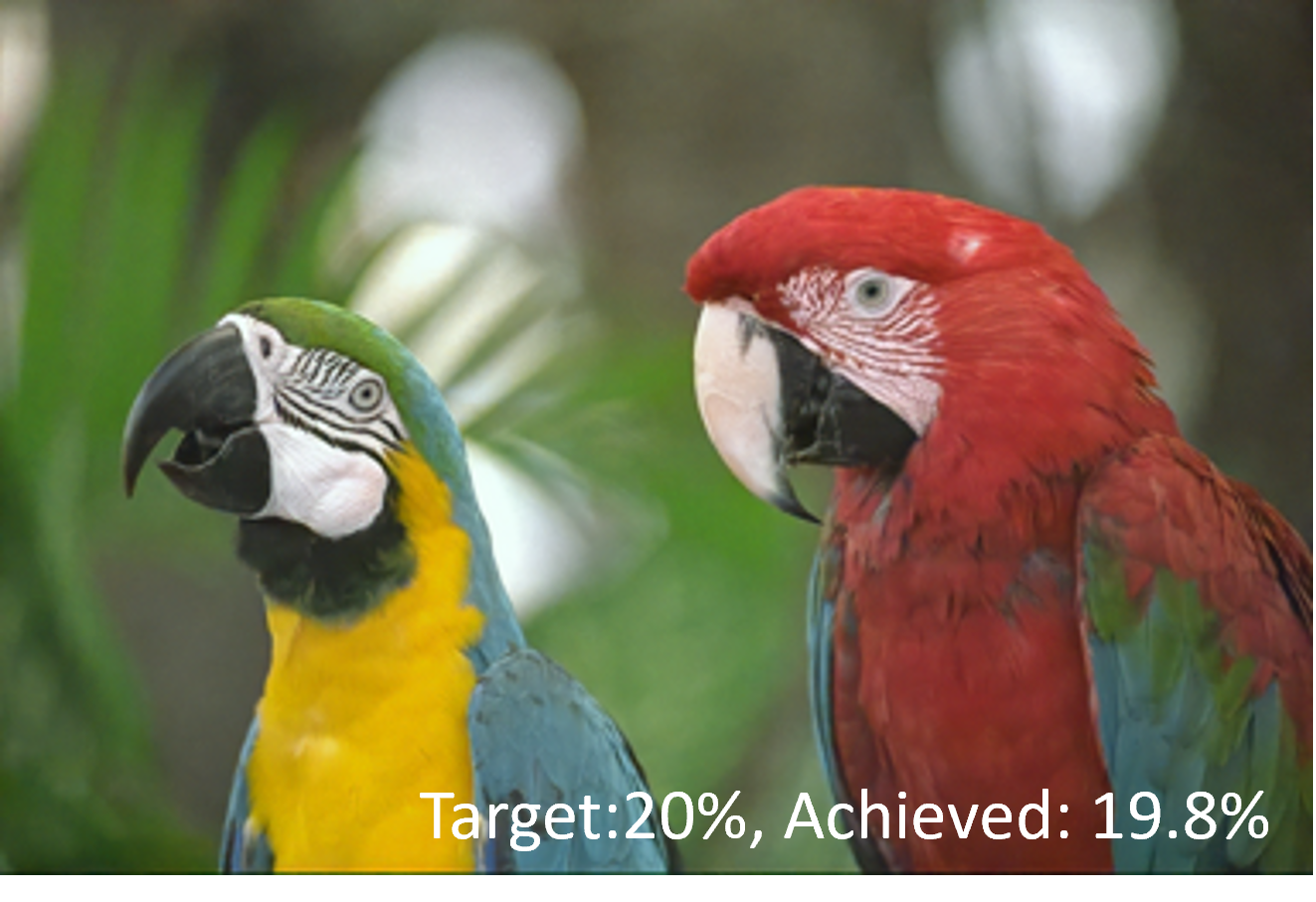}
    \end{subfigure}

    \begin{subfigure}[]{0.15\linewidth}
        \includegraphics[width=\linewidth]{figures/energy_lr/real/kodim23_combinaison_211_LR_ref.png}
                \caption*{Original}
    \end{subfigure}
    \begin{subfigure}[]{0.15\linewidth}
        \includegraphics[width=\linewidth]{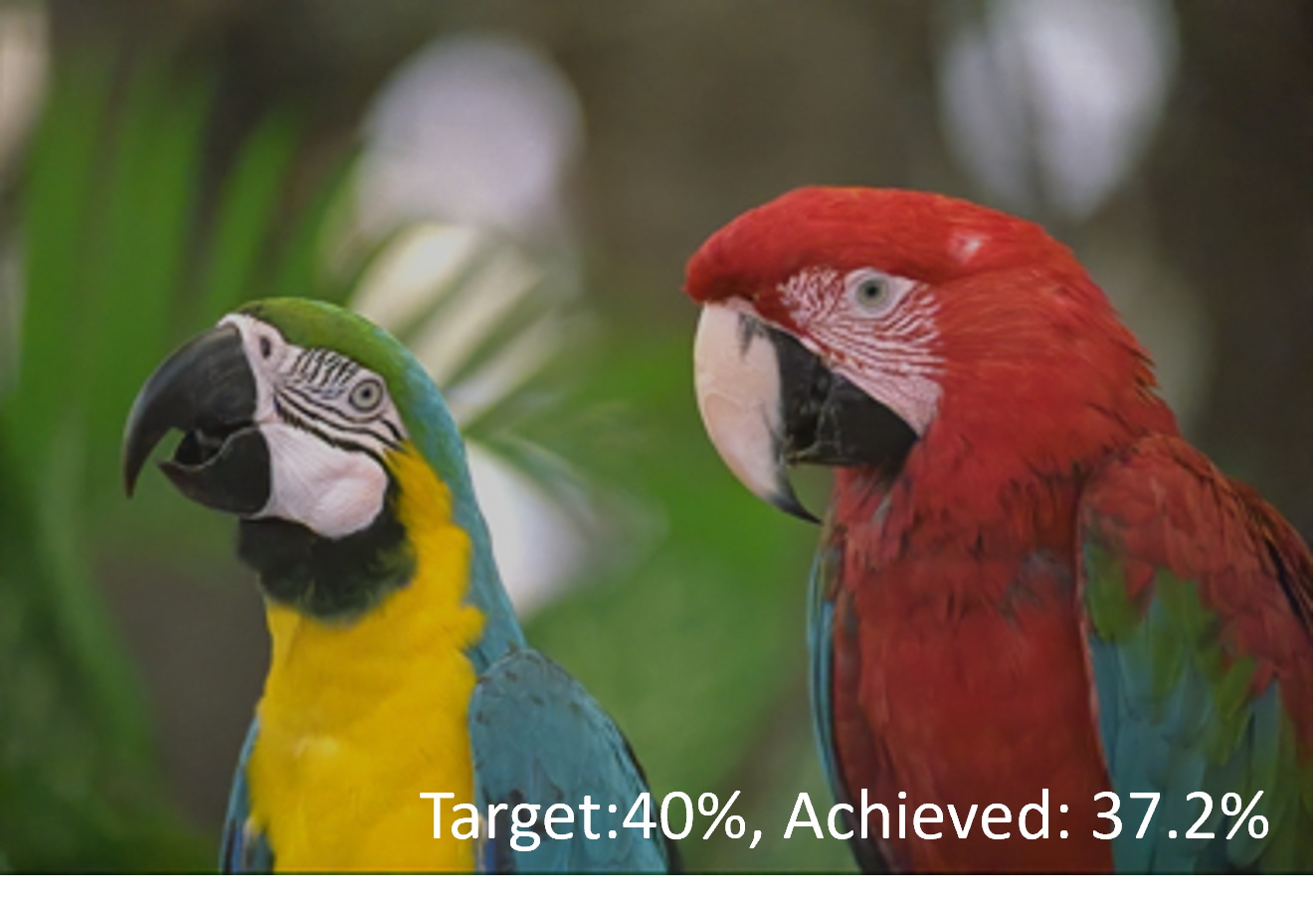}
         \caption*{LS}
    \end{subfigure}
    \begin{subfigure}[]{0.15\linewidth}
        \includegraphics[width=\linewidth]{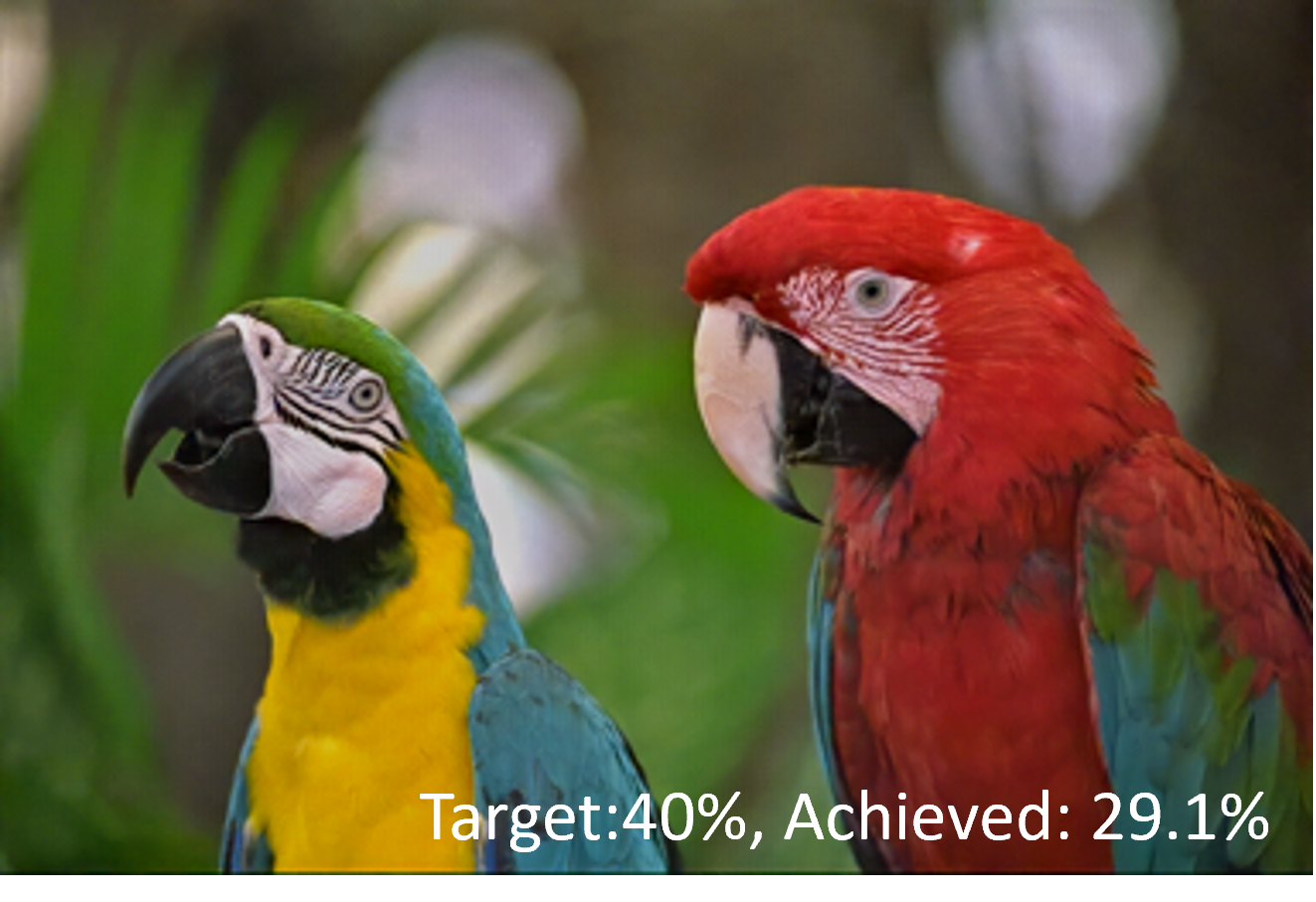}
        \caption*{RACE}
    \end{subfigure}
    \begin{subfigure}[]{0.15\linewidth}
        \includegraphics[width=\linewidth]{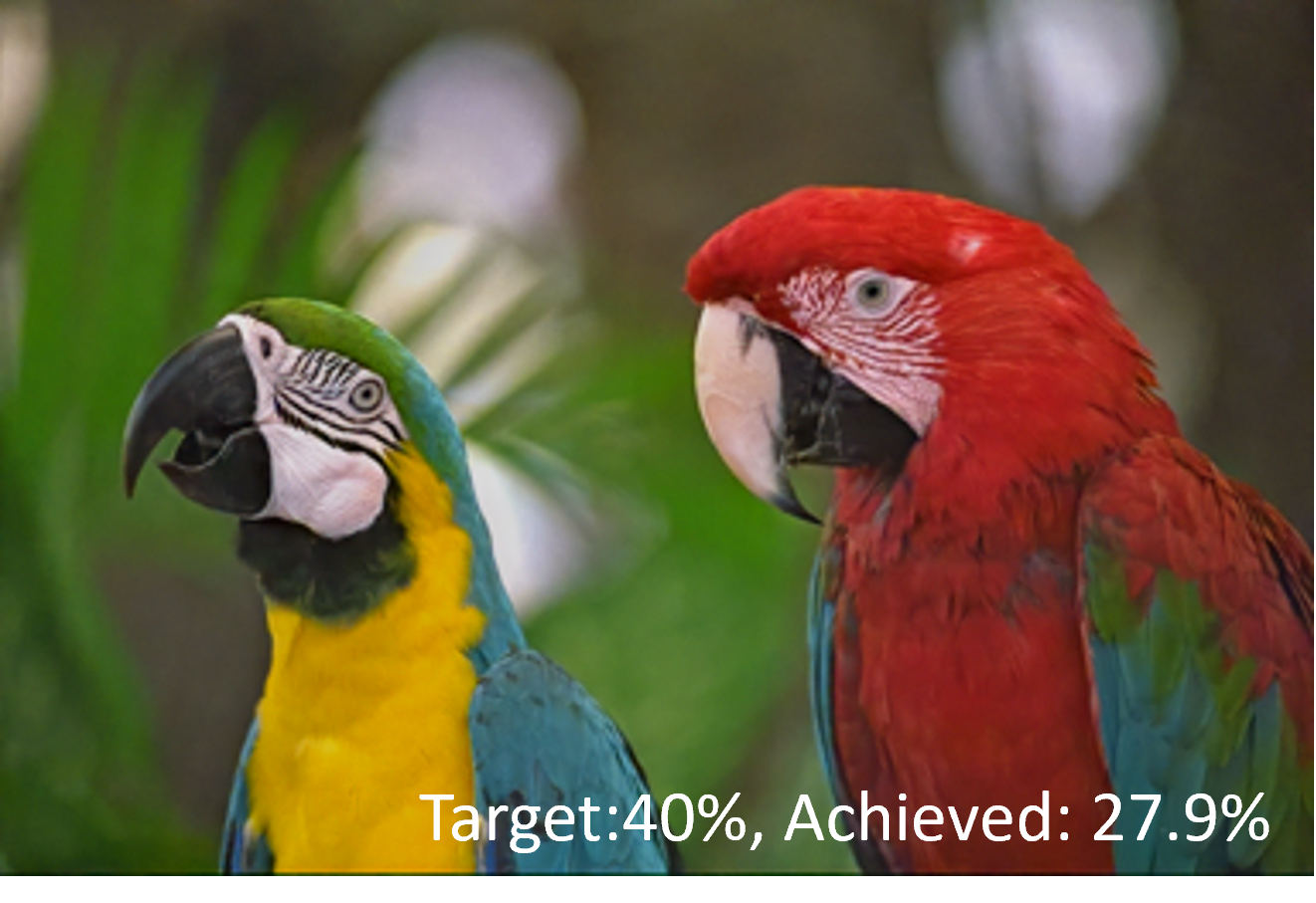}
        \caption*{DeepPVR}
    \end{subfigure}
    \begin{subfigure}[]{0.15\linewidth}
        \includegraphics[width=\linewidth]{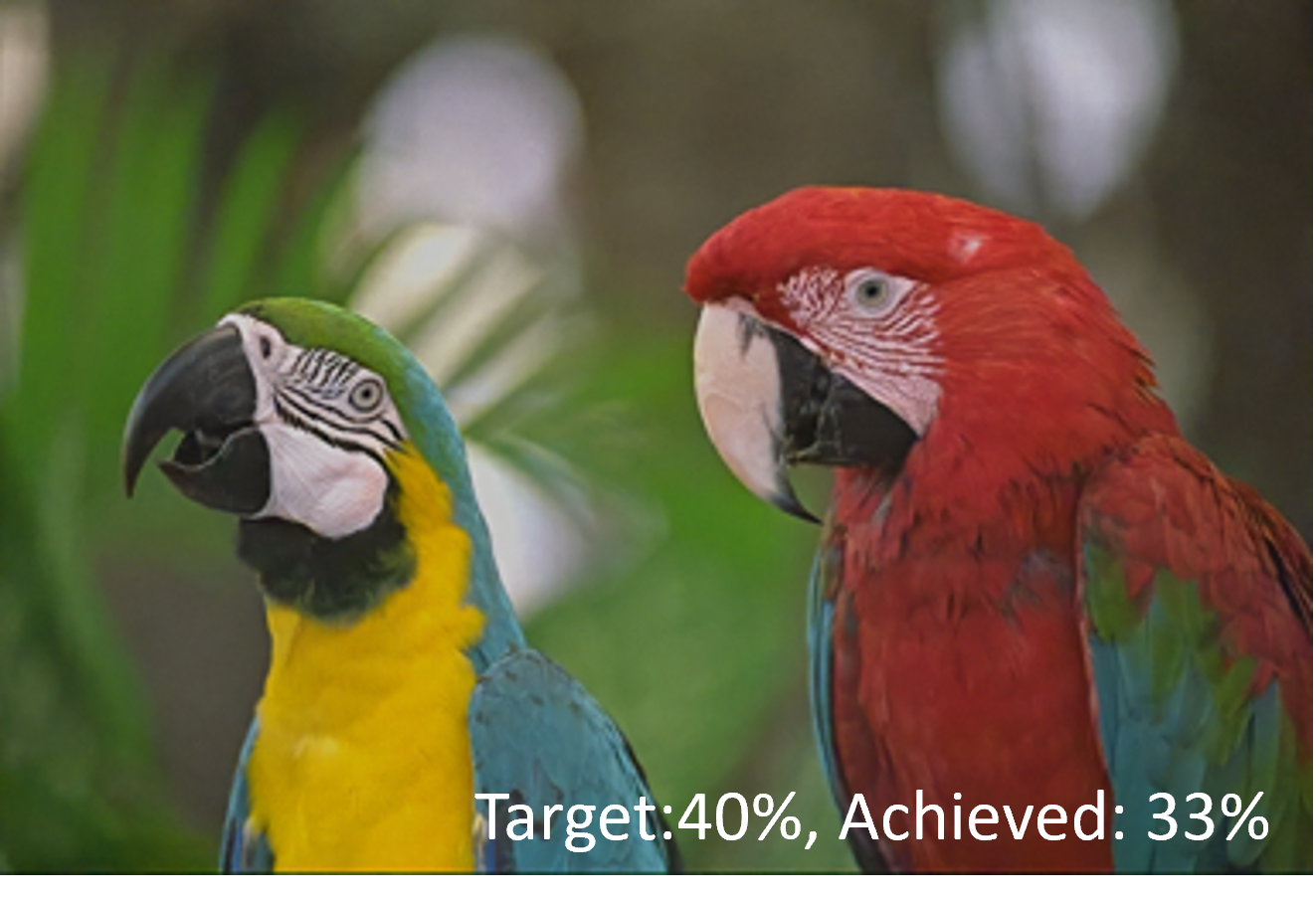}
        \caption*{\ac{invean}}
    \end{subfigure}
    \begin{subfigure}[]{0.15\linewidth}
        \includegraphics[width=\linewidth]{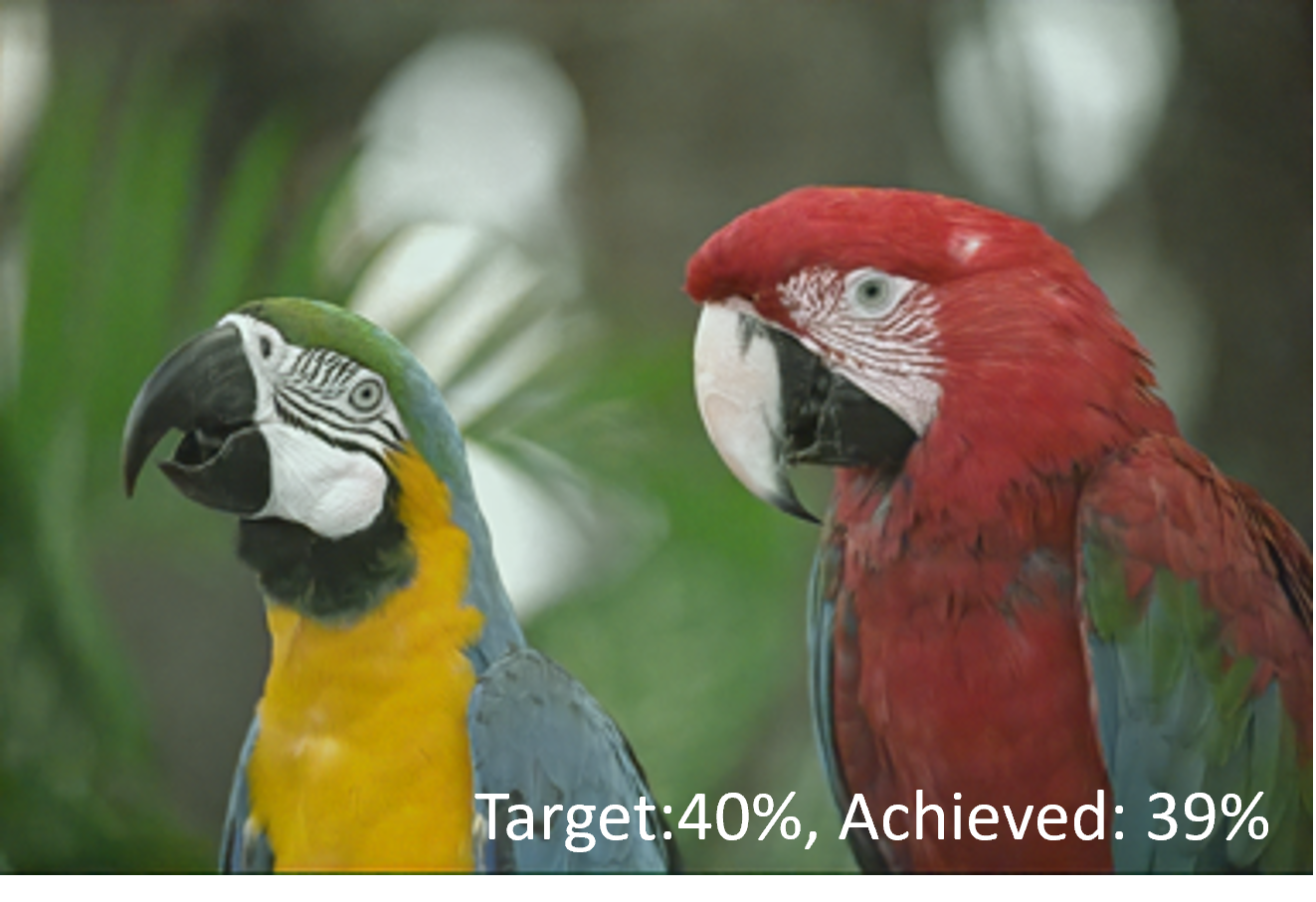}
        \caption*{Ours}
    \end{subfigure}
\caption{Comparison of generated energy-aware images with the state-of-the-art, for $R \in \{5\%, 20\%, 40\%\}$ from first to third lines. Achieved rates computed by the power model in~\cite{Demarty2023} are provided.}
\label{fig:energy_comparison}
\end{figure*}


In the following, we assess the performances of 3R-INN in terms of quality of the downscaled grain-free energy-aware LR image, and of the reconstructed HR grainy and clean images, against state-of-art methods for the rescaling, film grain removal and synthesis, and energy-aware tasks.
\subsection{Evaluation of downscaled LR images}
\label{subsec:lr}
The quantitative and qualitative evaluation of the \ac{lr} clean image $\Tilde{I}_{LR|R=0}$, \ie, corresponding to an energy reduction rate $R=0$, is 
given in Table~\ref{tab:luma_lr_all} and Figure~\ref{fig:removal_lr}, respectively.
The reference image is the bicubic rescaling of the HR clean image. Although quite similar  to the experimental protocol used in~\cite{xiao2020invertible}, we here assess the ability of the network both to rescale and to remove film grain, since input images are grainy. To compare our results to those of the IRN method~\cite{xiao2020invertible} in a fair manner, we retrain IRN with our training set and with the loss functions used in~\cite{liu2021invertible}, for both rescaling and film grain removal. As InvDN~\cite{liu2021invertible} outputs a rescaled image with a factor higher than 2, it was not included in the comparison. 
Results show that the proposed method performs better than IRN in terms of PSNR and SSIM. They also outline the good generalization of the proposed method, as even better performances are observed on BSDS300 and Kodak24 datasets.

\begin{figure}[h!]
    \centering
        \includegraphics[width=\linewidth]{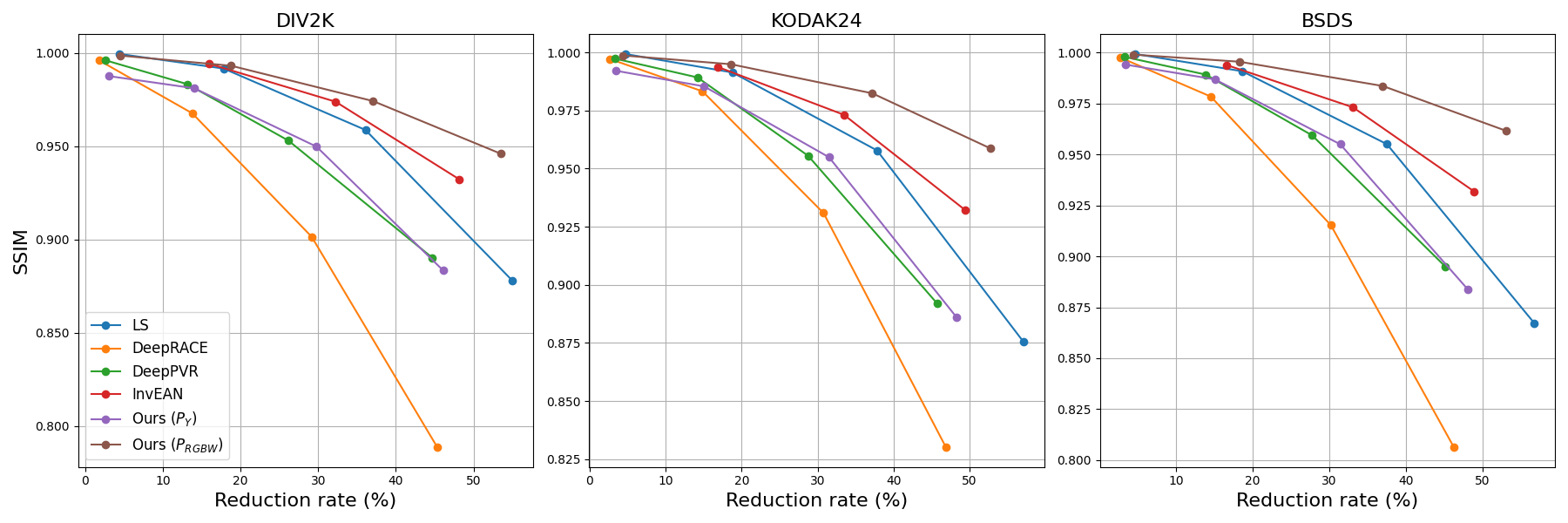}
\caption{SSIM scores as function of the target power reduction, for the different energy-aware methods.}
\label{fig:plots}
\end{figure}

For $R > 0$, we evaluate the visual quality of $\Tilde{I}_{LR|R}$ against state-of-the-art energy-aware methods, \ie, a global linear scaling of the luminance (LS), R-ACE~\cite{Nugroho2022r}, DeepPVR~\cite{LeMeur2023deep} and \ac{invean}~\cite{LeMeur2023invertible}. To solely evaluate the energy-aware task, and for a fair comparison, existing methods were evaluated while taking as input the output of our method after the fine tuning step with $R=0$. All evaluations metrics in the following were calculated with this image as reference.  
Table~\ref{tab:luma_energy_comparison} reports PSNR-Y and SSIM metrics at 4 reduction rates, on the three test sets. 
Two conclusions can be drawn. First, when the power consumption model $P_Y$ is used for a fair comparison with state-of-the-art methods, the proposed method outperforms LS and R-ACE methods, while being similar to DeepPVR and slightly below InvEAN. When the power consumption model $P_{RGBW}$ is used, the quality scores of the proposed method are significantly better, and especially for the PSNR-Y. This can be explained by the fact that our model does not learn to reduce the image luminance, contrary to state-of-art methods. The latter in turn were not trained to optimize  $P_{RGBW}$; this may explain their lower performances. 
This trend is confirmed by Figure~\ref{fig:plots} which plots \ac{ssim} scores as function of the actual reduction rate, computed with $P_{RGBW}$.  PSNR plots are provided in the supplemental material. 
Figure~\ref{fig:energy_comparison} shows a qualitative comparison of energy-aware images. 3R-INN and LS respect the reduction rate targets better than other methods. Our method also exhibits a different behavior for high values of $R$, once again keeping the luminance but modifying the colors. The subjective comparison is however difficult since the achieved energy reduction varies from one method to another.

In conclusion, 3R-INN, although not fully dedicated to the energy-reduction task, performs well compared to  existing methods. Additionally, similarly to \ac{invean}, the original image can be recovered without any side-information.

\begin{figure*}[t!]
    \centering
    \begin{subfigure}[]{0.19\linewidth}
        \includegraphics[width=\linewidth]{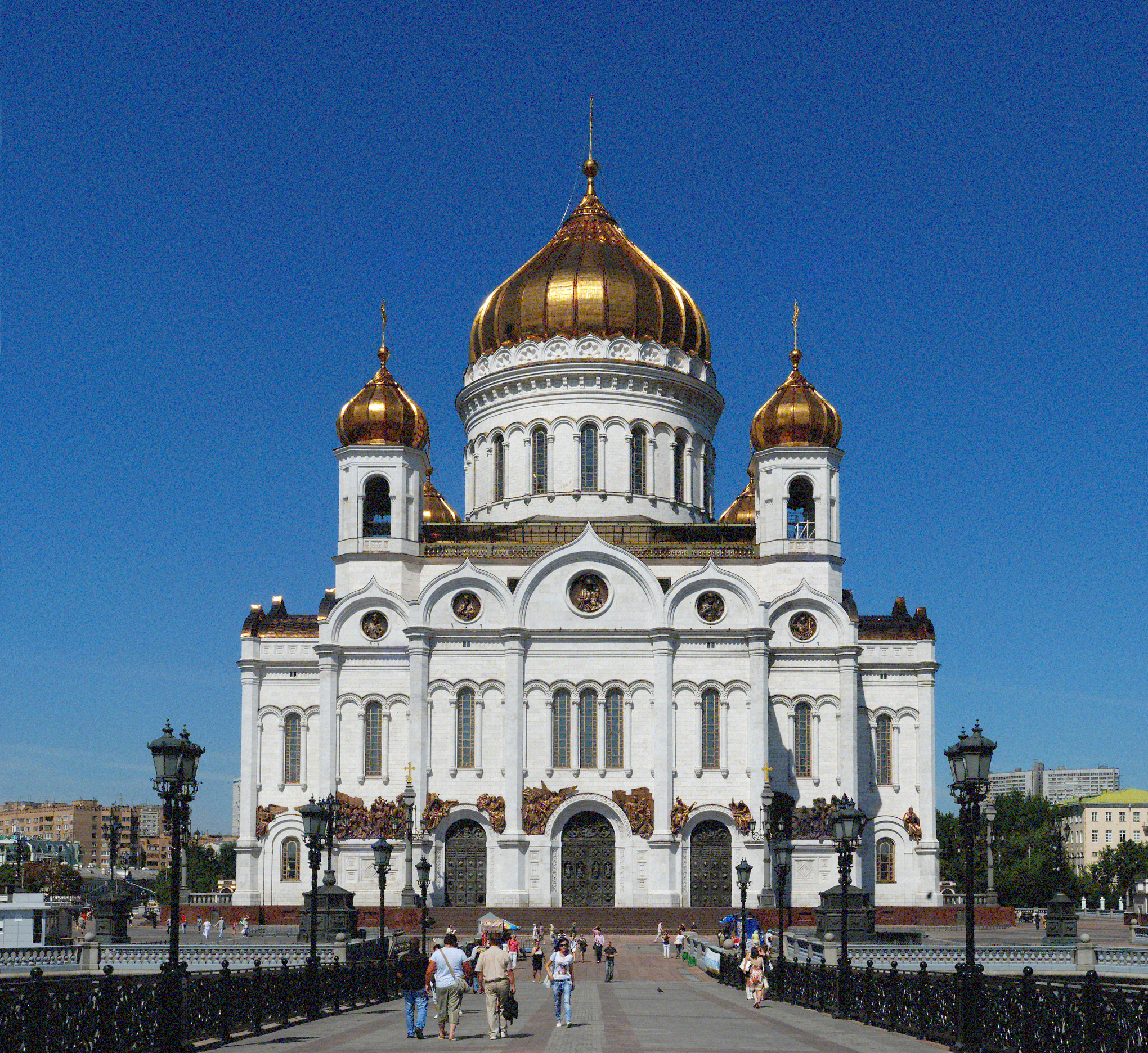}
        \caption*{Ground-truth}
    \end{subfigure}
    \begin{subfigure}[]{0.19\linewidth}
        \includegraphics[width=\linewidth]{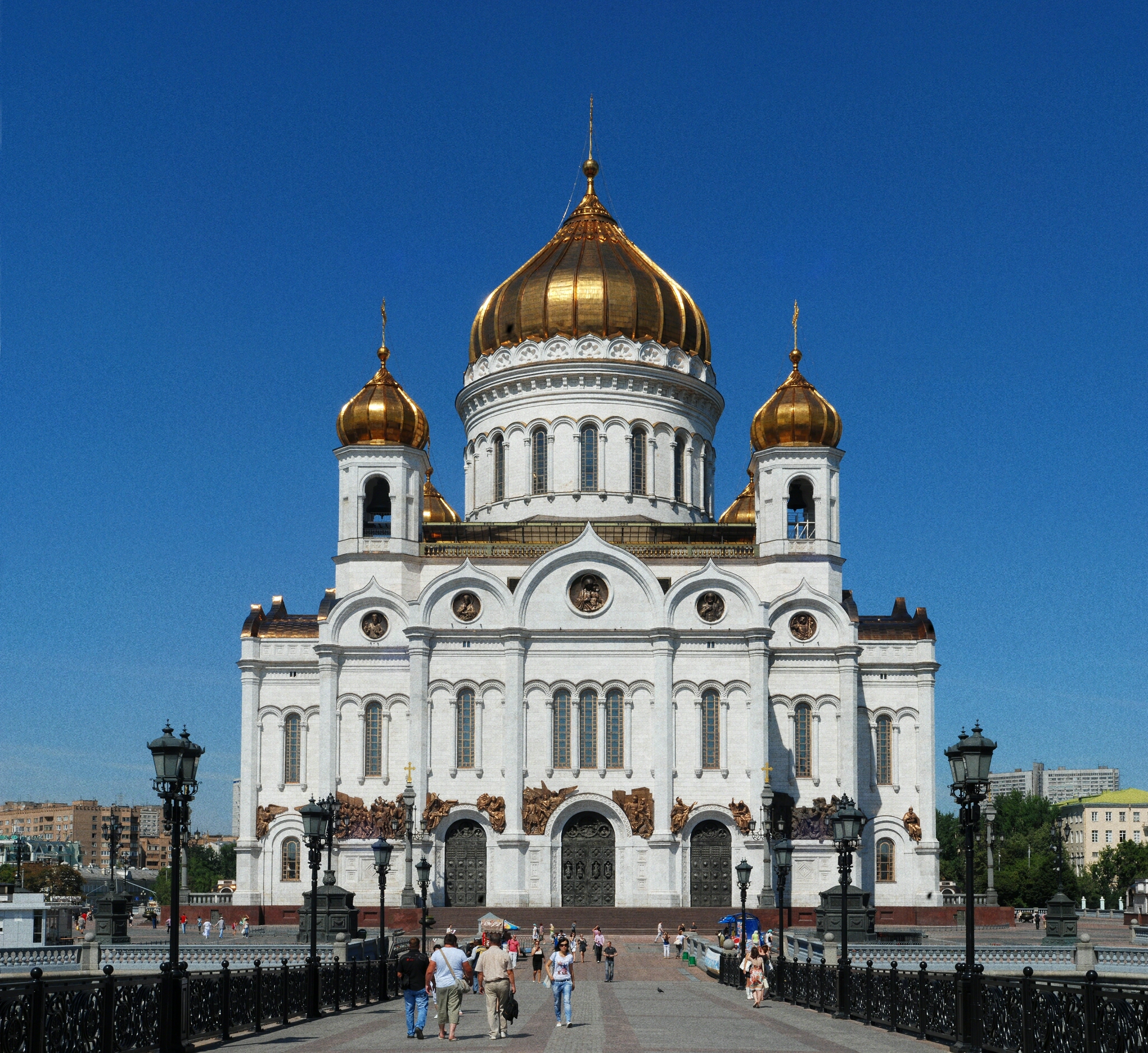}
        \caption*{VVC (LPIPS=0.3343)}
    \end{subfigure}
    \begin{subfigure}[]{0.19\linewidth}
        \includegraphics[width=\linewidth]{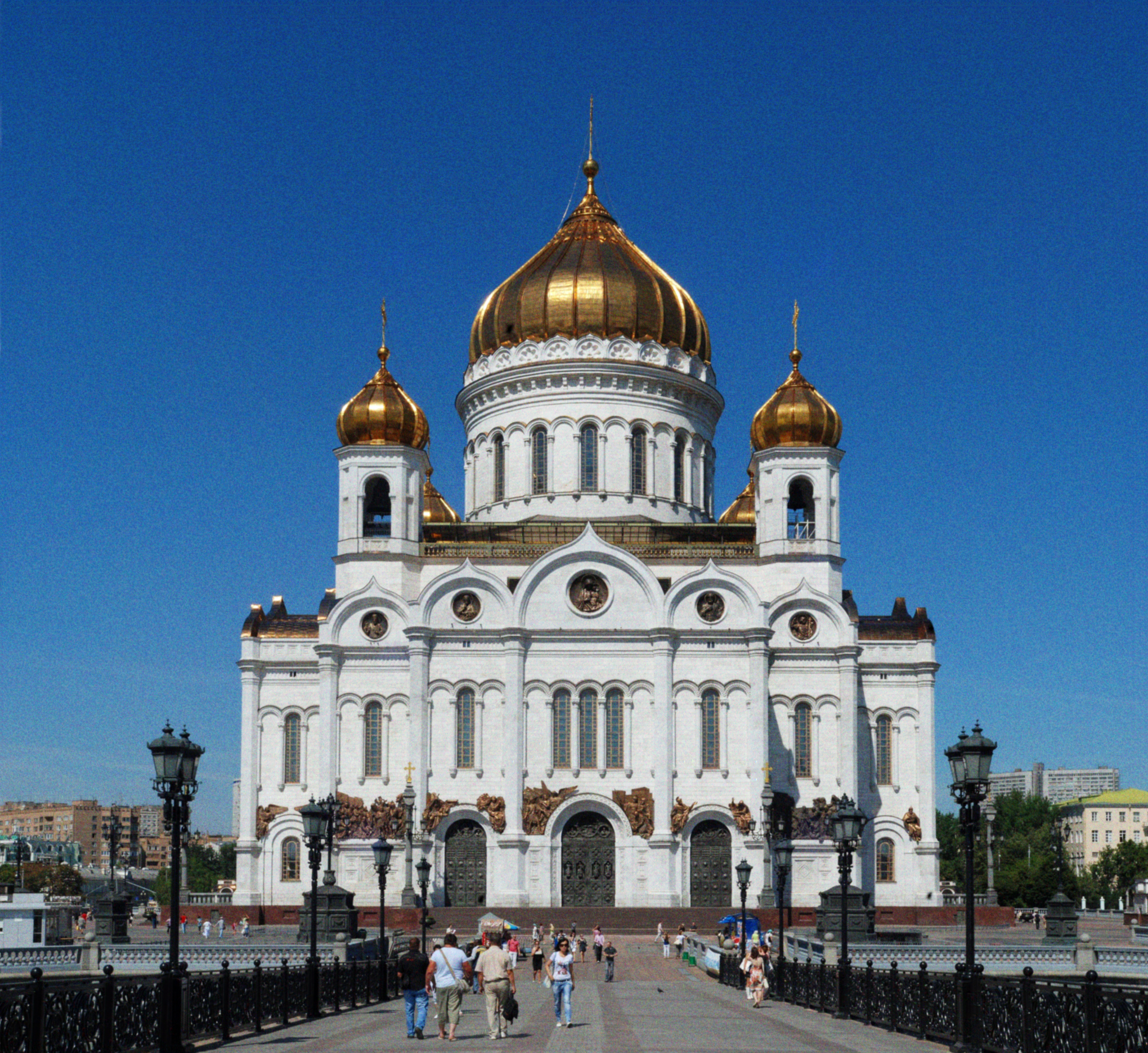}
        \caption*{DeepFG (LPIPS=0.3533)}
    \end{subfigure}
    \begin{subfigure}[]{0.19\linewidth}
        \includegraphics[width=\linewidth]{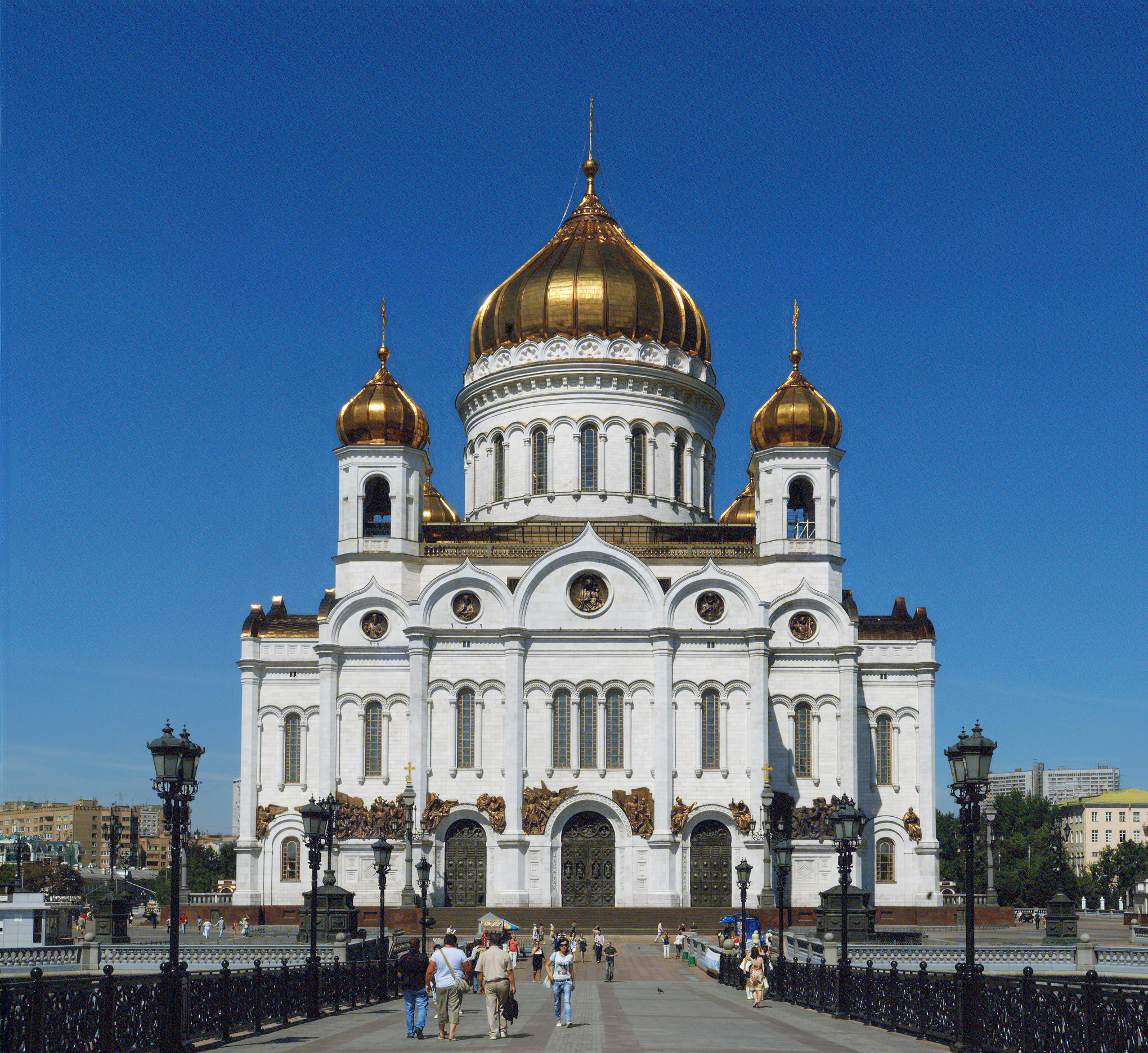}
        \caption*{StyleFG (LPIPS=0.1693)}
    \end{subfigure}
    \begin{subfigure}[]{0.19\linewidth}
        \includegraphics[width=\linewidth]{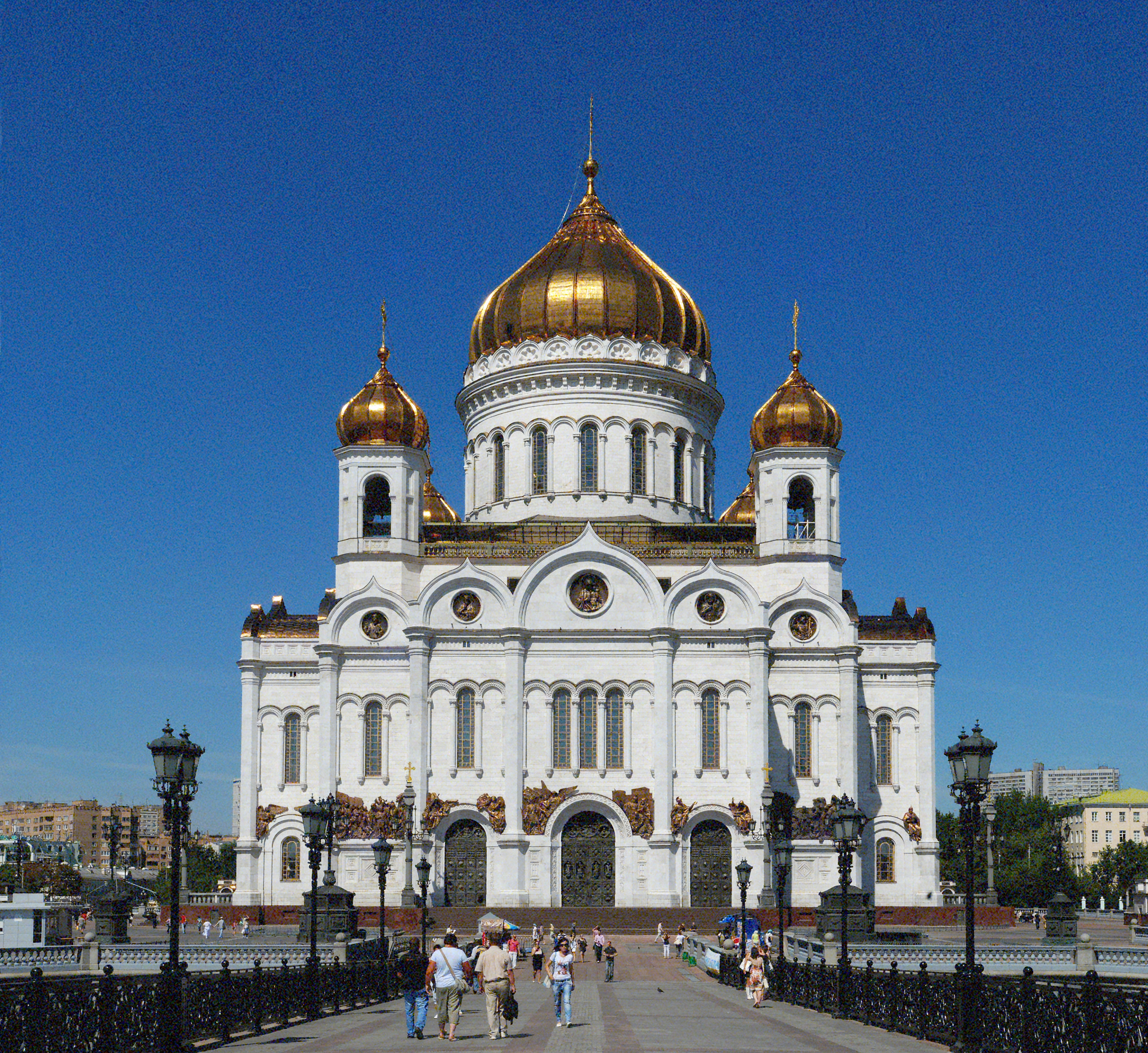}
        \caption*{Ours (LPIPS=0.0508)}
    \end{subfigure}
\caption{Qualitative evaluation of \ac{hr} synthesized grainy images for different methods, with LPIPS values.}
\label{fig:synthesis_comparison}
\end{figure*}

%
%
\subsection{Evaluation of generated HR clean images}
\begin{figure*}[]
    \centering

    \begin{subfigure}[]{0.24\linewidth}
        \includegraphics[width=\linewidth]{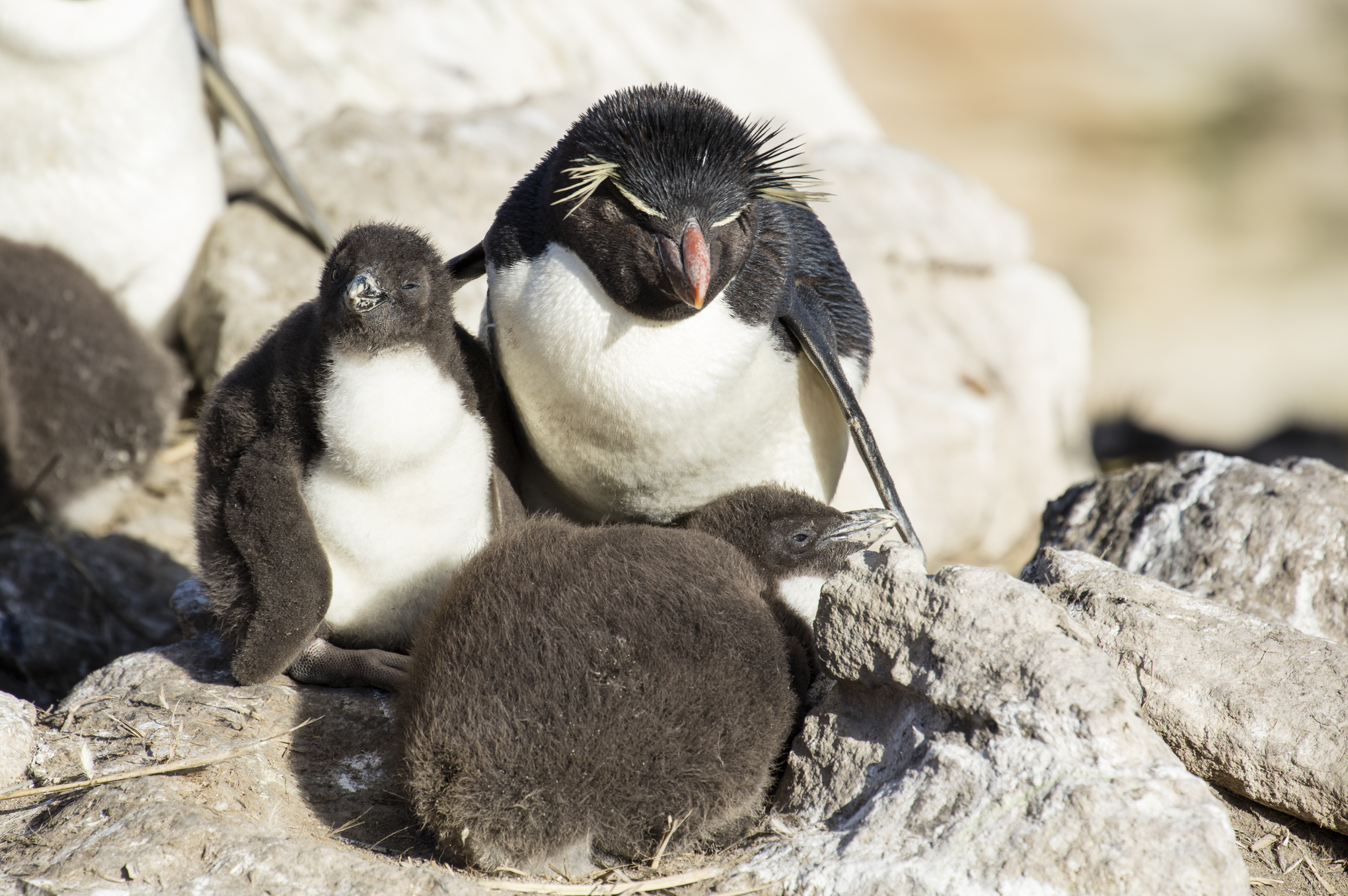}
        \caption*{Ground-truth clean}
    \end{subfigure}
    \begin{subfigure}[]{0.24\linewidth}
        \includegraphics[width=\linewidth]{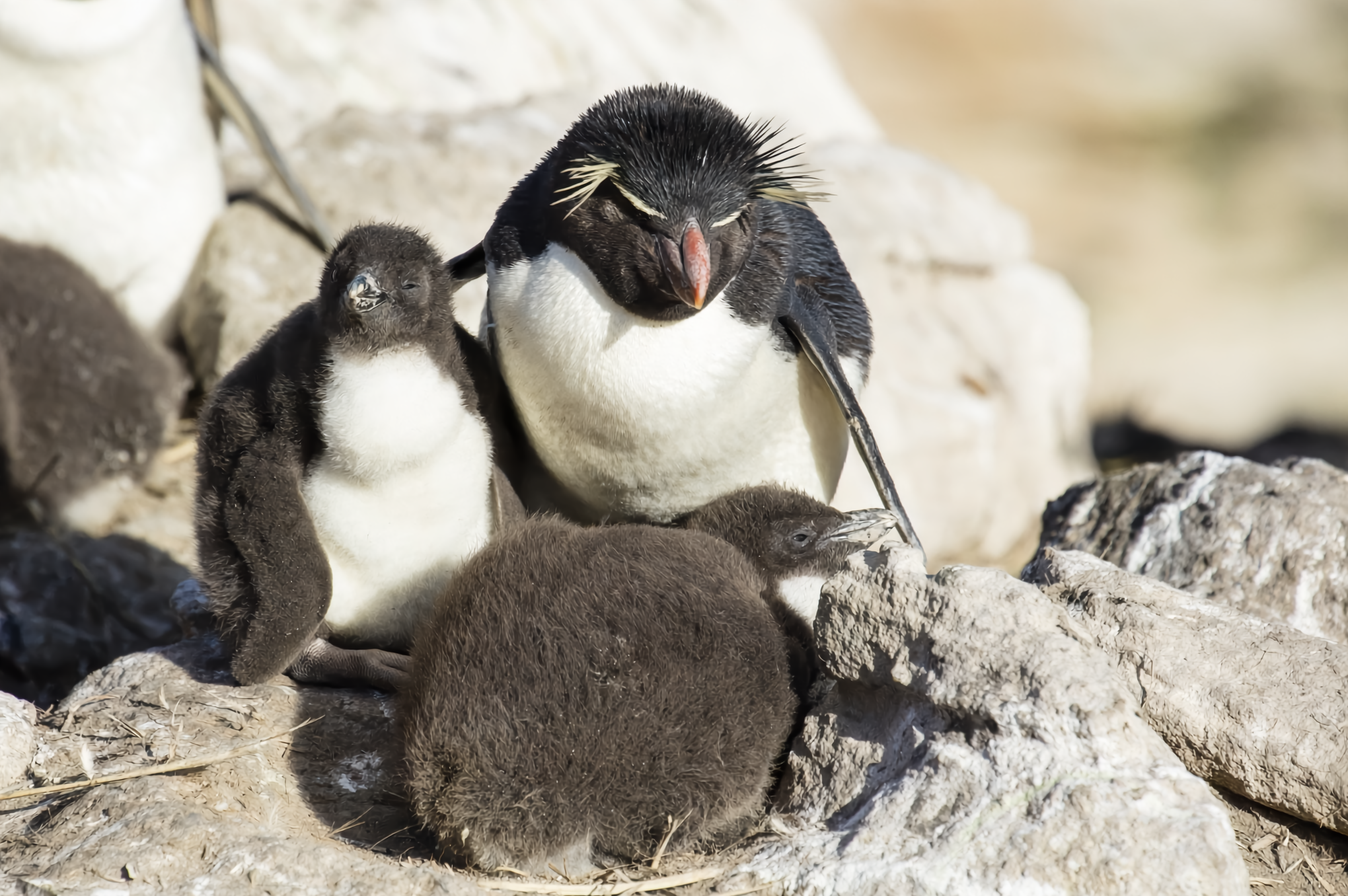}
        \caption*{InvDN (34.43/0.945)}
    \end{subfigure}
    \begin{subfigure}[]{0.24\linewidth}
        \includegraphics[width=\linewidth]{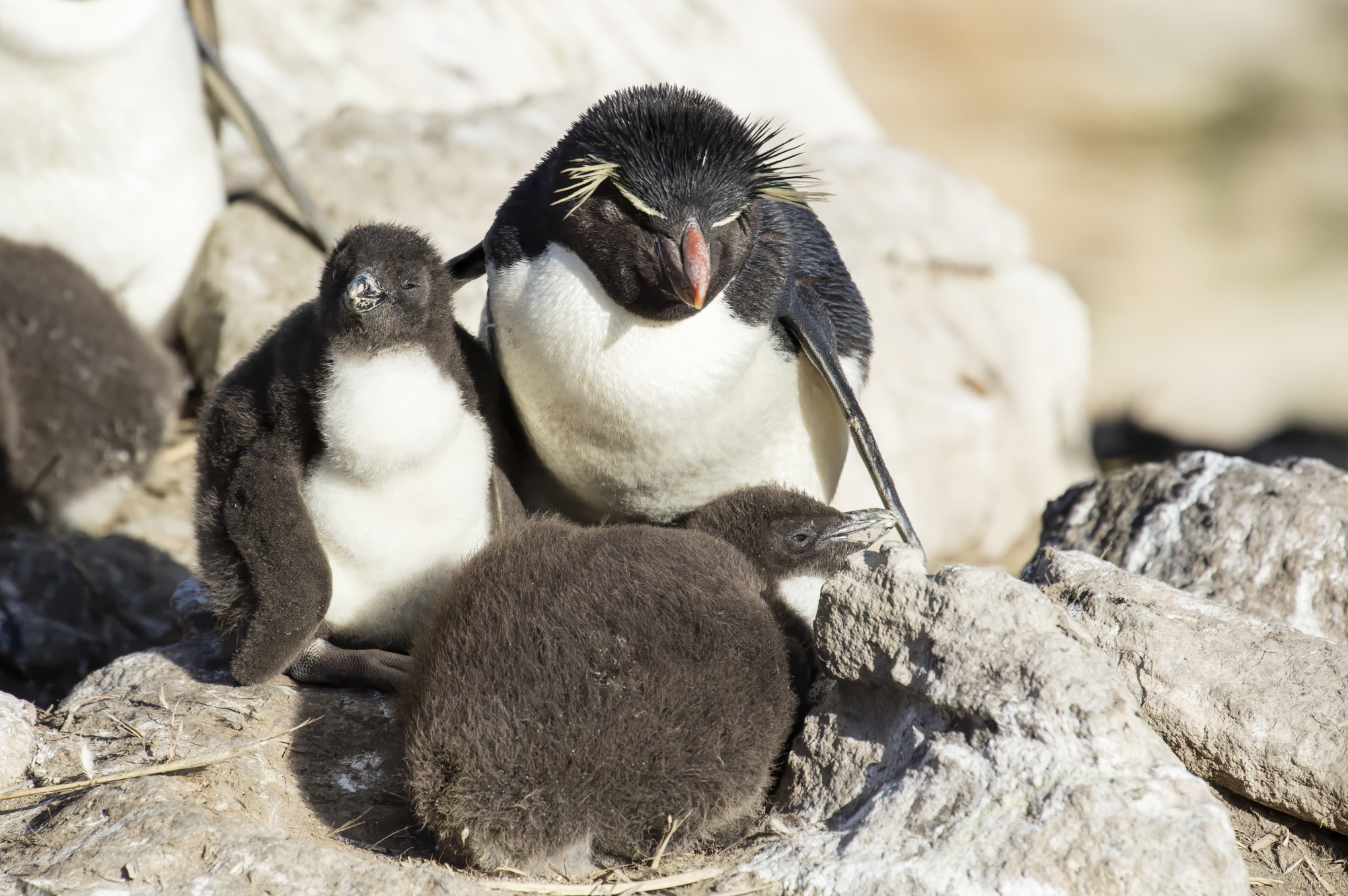}
        \caption*{IRN (39.97/0.975)}
    \end{subfigure}
    \begin{subfigure}[]{0.24\linewidth}
        \includegraphics[width=\linewidth]{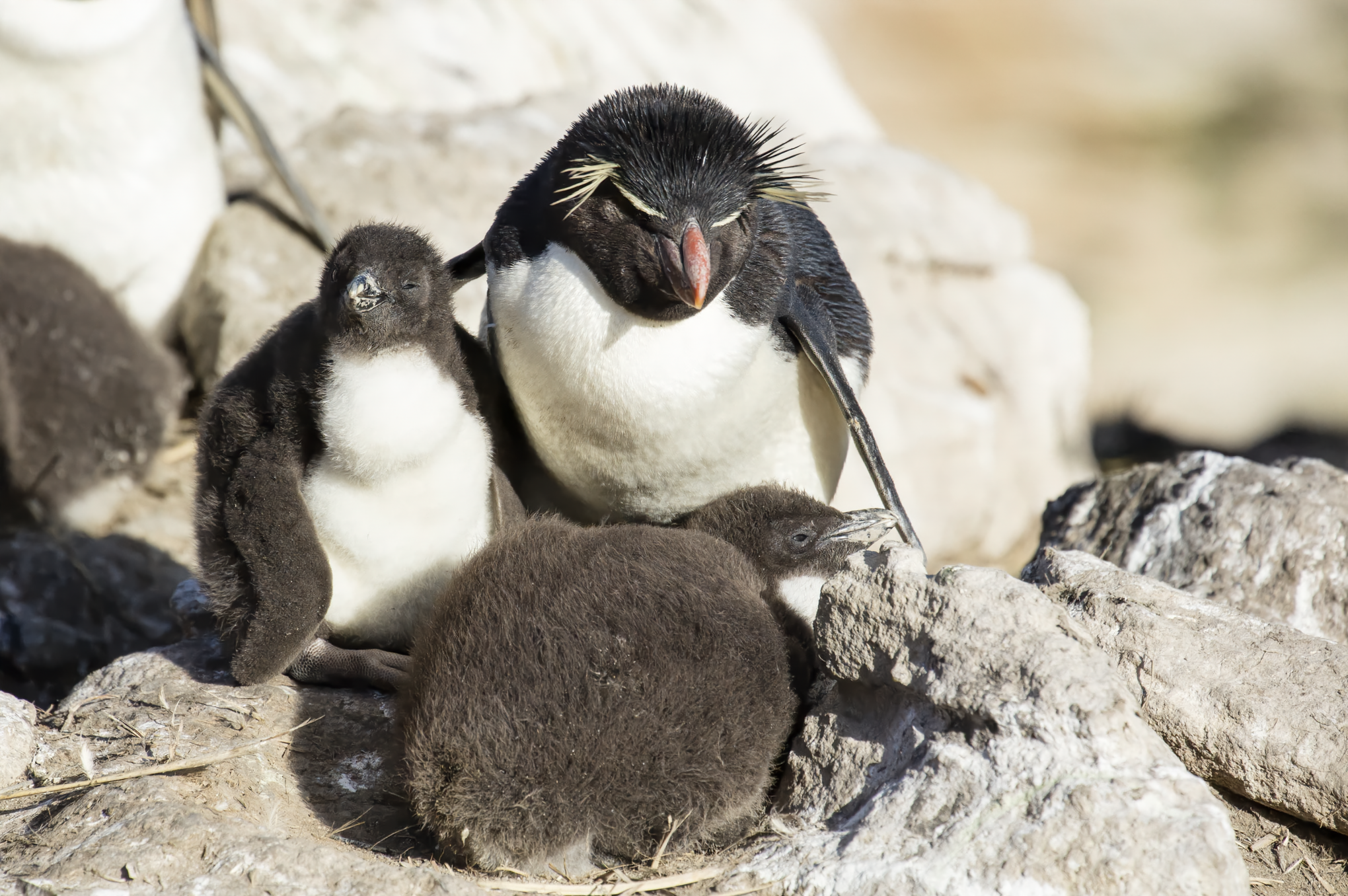}
        \caption*{Ours (38.53/0.969)}
    \end{subfigure}
\caption{Qualitative comparison between the reconstructed clean \ac{hr} images by InvDN, IRN and our model (PSNR/SSIM).}
\label{fig:clean_synthesis_comparison}
\end{figure*}

Another benefit of the proposed method is its ability to restore a \ac{hr} clean image.
Table~\ref{tab:luma_hr_all} presents a comparison in terms of PSNR-Y and SSIM, with IRN~\cite{xiao2020invertible} and \ac{invdn}~\cite{liu2021invertible} methods, re-trained as explained in section~\ref{subsec:lr}, for a fair comparison. 
Results indicate that the proposed method significantly outperforms \ac{invdn}~\cite{liu2021invertible}. Compared to IRN~\cite{xiao2020invertible}, we observe a significant difference in terms of PSNR (in average 1.3 dB) and a slight difference in terms of SSIM (in average 0.01). 
A qualitative evaluation is also proposed in Figure~\ref{fig:clean_synthesis_comparison}. The reconstructed clean \ac{hr} images show comparable quality for both our model and IRN.
\begin{table}[]
\caption{Comparison between reconstructed \ac{hr} clean images and ground-truth in terms of PSNR and \ac{ssim}.}
\begin{adjustbox}{max width=\linewidth}
\begin{tabular}{l|c|cc|cc|cc}
\toprule
\multirow{2}{*}{Method} & \multirow{2}{*}{Nb parameters} & \multicolumn{2}{c}{DIV2K} & \multicolumn{2}{c}{BSDS300} & \multicolumn{2}{c}{Kodak24} \\
                         &  & PSNR $\uparrow$ & SSIM $\uparrow$ & PSNR $\uparrow$ & SSIM $\uparrow$ & PSNR $\uparrow$ & SSIM $\uparrow$\\\midrule
IRN \cite{xiao2020invertible}   &1.66M       &\textbf{36.53} &\textbf{0.927}         &\textbf{35.22} &\textbf{0.939}   &\textbf{36.21} &\textbf{0.935}        \\
InvDN \cite{liu2021invertible}  &2.64M        &33.15 &0.891      &26.50 &0.787 &31.99  &0.880                                         \\   
Ours &1.74M &35.43 &0.915 &33.86 &0.923 &34.83 &0.917
\\ \bottomrule          
\end{tabular}
\end{adjustbox}
\label{tab:luma_hr_all}
\end{table}
%
\subsection{Evaluation of reconstructed HR grainy images}
One important feature of 3R-INN is its reversibility property. To evaluate this property, we compared the performance of the \ac{hr} grainy image reconstruction with state-of-the-art film grain synthesis methods, \ie, VVC (Versatile Video Coding) implementation~\cite{radosavljevic2021implementation}, Deep-FG~\cite{ameur2023deep} and Style-FG~\cite{ameur2023deep}. As Deep-FG does not do any analysis of the grain, for a fair comparison, we generate 5 versions of film grain, one per available intensity level, and kept only the best performing image for each metric in the comparison.
Table~\ref{tab:synthesis_comparison} summarizes the quantitative results for 3R-INN for $R = 0$, in terms of fidelity of the synthesized grain using \ac{lpips}, JSD-NSS and the \ac{kld}~\cite{zhu2016blind}, these last two being computed between the histograms of ground-truth and \ac{hr} grainy images.
Similar results are obtained for $R > 0$ and are presented as supplemental material. 
\begin{table}[]
\caption{Comparison between reconstructed \ac{hr} grainy images and ground-truth in terms of \ac{jsdnss}, \ac{lpips} and \ac{kld} for different methods on DIV2K validation set.}
\centering
\begin{adjustbox}{max width=\linewidth}
\begin{tabular}{l|c|c|c|c|c|c}
\toprule
&Nb parameters &Analysis &Auxiliary data & JSD-NSS $\downarrow$ & LPIPS $\downarrow$ & KLD $\downarrow$ \\
\midrule
VVC\cite{radosavljevic2021implementation} &- &\checkmark &set of params & 0.0148 & 0.2981 & 0.0327 \\
Deep-FG \cite{ameur2023deep} &32M &x &x      & 0.0134 &0.3722 &0.0260 \\
Style-FG \cite{ameur2023style} &20M+33M  &\checkmark &style vector &\textbf{ 0.0024} & 0.1592 & 0.0232 \\
Ours & 1.7M &\checkmark &none &0.0088 &\textbf{0.0445} &\textbf{0.0177} 

\\
\bottomrule  
\end{tabular}
\label{tab:synthesis_comparison} 
\end{adjustbox}
\end{table}
Results show that the proposed method outperforms quantitatively VVC~\cite{radosavljevic2021implementation}, Deep-FG~\cite{ameur2023deep}. It also performs better than Style-FG~\cite{ameur2023deep} for LPIPS and KLD metrics which are representative of the quality of generated grain. The lower JSD-NSS value for Style-FG~\cite{ameur2023deep} could be explained by the fact that it is a GAN network, which therefore tries to first model the distribution of the data, at the expense of the output quality. These observations are confirmed by the qualitative comparison, as illustrated by Figure~\ref{fig:synthesis_comparison} (additional results in the supplemental material). 
As additional advantage, the proposed method  does not need to transmit auxiliary data for synthesizing grain.
%
%
\subsection{Ablation study}
We investigated 1) the benefit of using the latent encoding block, 2) the weighting of the losses, and 3) the size of the disentanglement. Table \ref{tab:z_ablation} shows results for incremental versions of our model, on the DIV2K validation set. 

\noindent \textbf{Latent encoding block }
We investigated three configurations to capture the lost information $z$ in the forward pass and to reconstruct both $\Tilde{I}_{C}$ and $\Tilde{I}_{G}$ in the inverse pass. \textit{Config.1} restores $\Tilde{I}_{C}$ using a random Gaussian distribution sample, and $\Tilde{I}_{G}$ using the original high-frequency signal. It achieves the best reconstructions, but at the expense of transmitting $z$. \textit{Config.1} corresponds to the upper-bound quality we could reach. For the sake of operational implementation and energy savings, we want to avoid the transmission of $z$. A baseline configuration \textit{Config.2} therefore consists in reconstructing both $\Tilde{I}_{C}$ and $\Tilde{I}_{G}$ using a disentangled random Gaussian distribution sample that separates high-frequency details and film grain (dim($\tilde{z}_G$)=2)). We observe an expected loss of quality, but with the advantage of not transmitting $z$. Our proposal \textit{Config.3} restores both $\Tilde{I}_{C}$ and $\Tilde{I}_{G}$ using a disentangled random Gaussian distribution sample whose mean and variance are conditioned on the \ac{lr} image thanks to the conditional latent encoding block (dim($\tilde{z_G}$)=2). Results show that it achieves comparable performance with \textit{Config.2} while reconstructing $\Tilde{I}_{C}$, and better fidelity while reconstructing $\Tilde{I}_{G}$. This shows the benefit of using a conditional latent encoding block, enabling image-adaptive reconstruction conditioned on the \ac{lr} image. 

\begin{table}[]
\centering
\caption{Comparison between different configurations of our model in terms of PSNR, SSIM, LPIPS and JSD-NSS.}
\begin{adjustbox}{max width=.9\linewidth}
\begin{tabular}{l|c|c|c}
\toprule
\multirow{2}{*}{Method}  & Clean LR       & Clean HR      & Grainy HR     \\ \cmidrule{2-4}
                         & \multicolumn{2}{c|}{PSNR$\uparrow$ / SSIM$\uparrow$} & JSD-NSS$\downarrow$ / LPIPS$\downarrow$ \\  \midrule
\textit{Config.1}                 &39.06/0.942                &36.53/0.927              &  0             \\
\textit{Config.2}                 &38.62/0.920                &35.53/0.913     &0.0096/0.0402             \\
\textit{Config.3}                &38.71/0.921                &35.53/0.913              &0.0090/0.0381               \\ \midrule
+$\lambda_{1}=40$     &39.30/0.936    &35.41/0.915  &0.0090/0.0381   \\\midrule

dim$(\tilde{z}_{G}) = 1$       &\textbf{39.45/0.937}   &\textbf{35.52}/0.914   &\textbf{0.0086}/0.0377          \\
dim$(\tilde{z}_{G}) = 2$       &39.30/0.936    &35.41/\textbf{0.915}  &0.0090/0.0381         \\
dim$(\tilde{z}_{G}) = 3$       &39.34/0.936    &35.37/0.914   &0.0087/\textbf{0.0366}          \\
dim$(\tilde{z}_{G}) = 4$       &39.37/0.937    &35.38/0.914  &0.0091/0.0390     \\

 \bottomrule      
\end{tabular}
\end{adjustbox}
\label{tab:z_ablation}
\end{table}

\noindent \textbf{Loss weighting}
In the previous experiments, $\lambda_1$ was set to 16. To further increase the quality of the clean \ac{lr}, we set its value to 40, adjusting the balance between the losses and letting the fidelity loss play a bigger role during training. 

\noindent \textbf{Disentangled representation}
We investigated varied dimensions of $\tilde{z_G}$. In general, extending dimensions assigned to film grain does not improve the film grain synthesis performance. On the other hand, it deteriorates both the quality of the clean \ac{hr} and \ac{lr} images. Thus, we use dim($\tilde{z}_G$) = 1 in our experiment.
A visualisation of the disentanglement of $\tilde{z}$ is provided in the supplemental material.
\subsection{End-to-end energy reduction}
The original goal of our paper is to reduce the overall energy consumption along the video distribution system. 

In that sense, 3R-INN performs three tasks and counts less than 2M parameters. This is to be compared with the NN-based post-processings implemented in JVET Neural Network-based Video Coding~\cite{galpin2023}, which tot up more than 5M for all three tasks, as super-resolution itself counts three networks of 1.5M each. The cost of re-running the framework to restore the original content is this time to be compared with the best existing deep learning-based methods for all three tasks: styleFG (53M) + IRN (1.66M) + InvEAN (806K), in total more than 55M parameters.

\begin{figure}[h!]
    \centering
        \includegraphics[width=\linewidth]{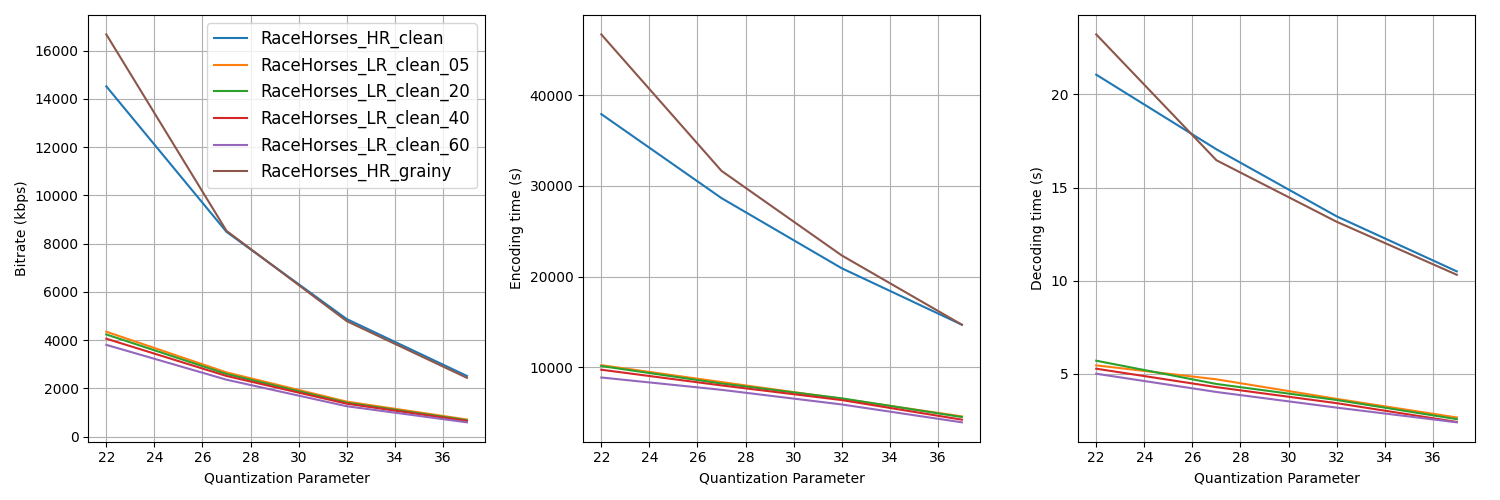}
\caption{Bit-rate, encoding and decoding times before and after using 3R-INN in terms of QP for sequence RaceHorses.}
\label{fig:time_bitrate_plots}
\end{figure}

We also tested the full video transmission chain by applying 3R-INN on two JVET sequences RaceHorses (300 frames, $832\times 480$), BasketBall (500 frames, HD)~\cite{boyce2018jvet}, encoding and decoding the LR outputs using VTM~\cite{VTM19}, in full intra mode, and re-applying 3R-INN in an inverse pass. Figure~\ref{fig:time_bitrate_plots} reports the average encoding/decoding times and bit-rates, for different QPs, for the HR clean and grainy RaceHorses sequences, and for LR versions with different $R \in [5\%; 20\%; 40\%; 60\%]$. Up to QP = 27, encoding and decoding the HR grainy video is more time and bit-rate demanding than for the HR clean version. For higher QPs, encoding time is still higher, however, bit-rate and decoding time are similar, because grain was removed during the encoding process. This confirms that compressing a grainy video while preserving film grain requires encoding at low QPs (which is far from the real-world scenario), leading to high and impractical bit-rates. On the contrary, encoding LR, grain-free versions, whatever the value of $R$, shows substantially lower times and bit-rates, and consequently reduces the energy at the head-end, transmission and decoding stages. These figures translate into 78\%, 3\% and \textit{ca.} 77\% of savings for respectively head-end, delivery and decoding, for the sequence RaceHorses, at QP22 and R = 20\%, according to the energy model described in~\cite{Malmodin2020power, Herglotz2023sweet}. Detailed computation is provided in the supplemental material.

Figure~\ref{fig:energy_measures} presents actual measures of energy consumptions for $R \in [5\%, 20\%, 40\%, 60\%]$, on an \ac{oled} LG-42C2 screen, for the sequence RaceHorses. On the left plot, we compare the consumption of the encoded/decoded LR and HR clean sequences at QP = 22. This proves that displaying an energy-aware video at different reduction rates significantly reduces the display power consumption, although some improvment still needs to be made to attain the expected target (average powers are: 6.8\%, 21.5\%, 33.3\%, 44.2\%). The right plot shows a comparison of the consumption of the LR sequences for different $R$, before and after encoding/decoding (QP = 22). For each $R$, both curves are rather similar and respect the same ordering. This proves that energy-aware images are to some extent robust to compression in terms of power values. Similar results are obtained for the sequence BasketBall (shown in the supplemental material).
\begin{figure}[h!]
    \centering
    \begin{subfigure}[]{0.45\linewidth}
        \includegraphics[width=\linewidth]{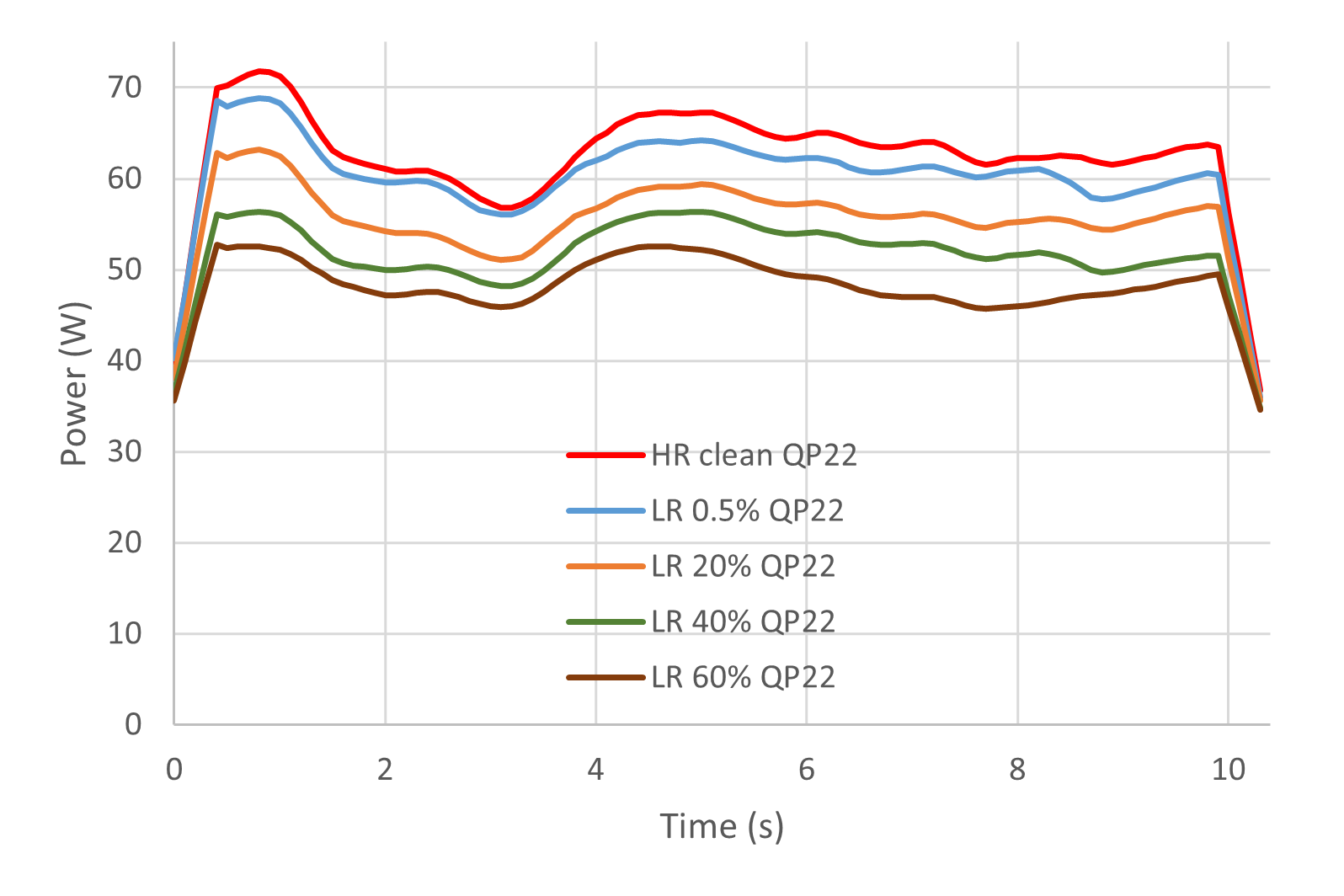}
         \caption*{}
     \end{subfigure}
      \begin{subfigure}[]{0.45\linewidth}
        \includegraphics[width=\linewidth]{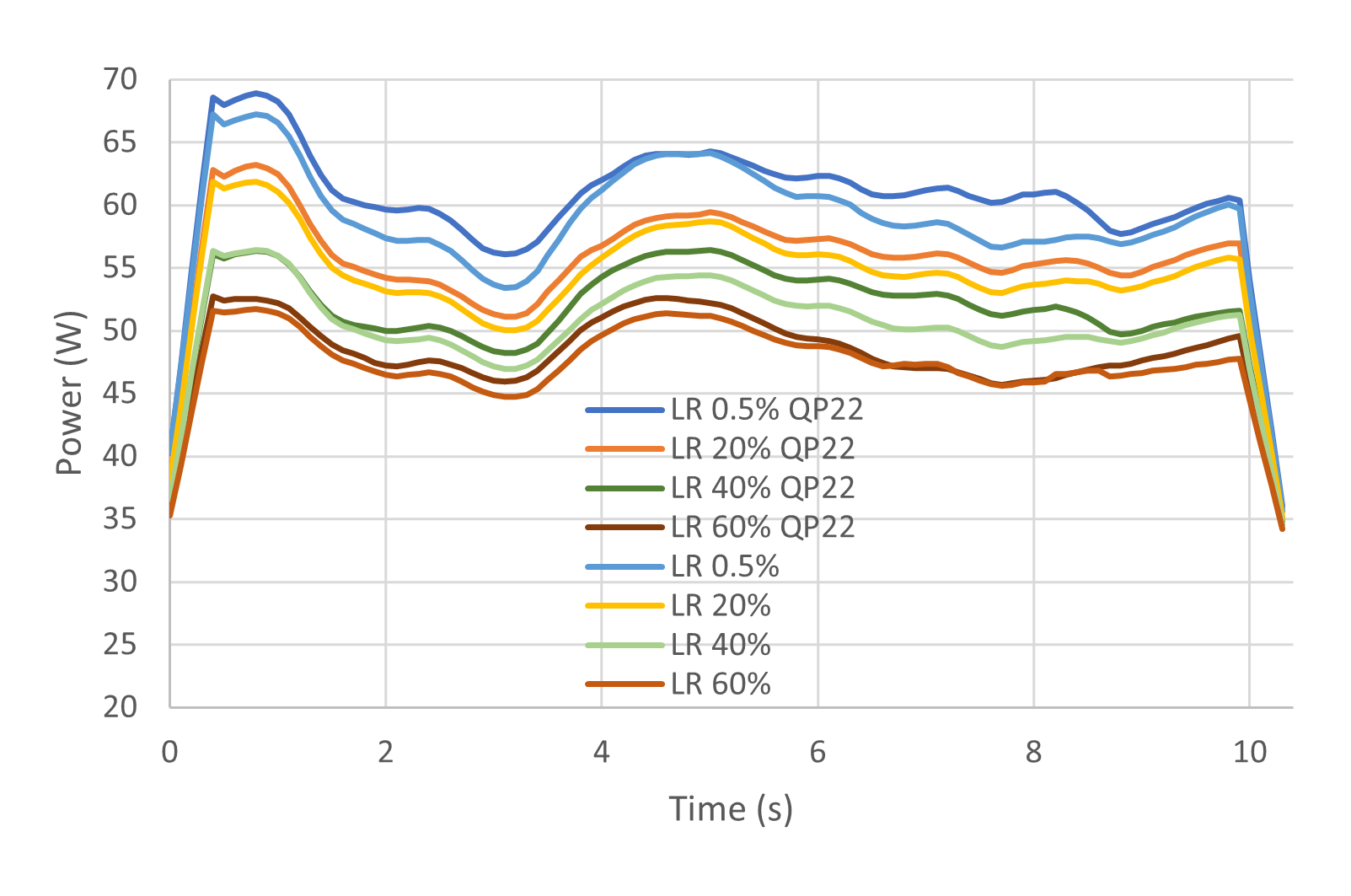}
         \caption*{}
     \end{subfigure} 
\caption{Measured power consumption for sequence RaceHorses. Left: Comparison between HR and LR versions at QP=22. Right: Comparison between LR versions before and after encoding/decoding.}
\label{fig:energy_measures}
\end{figure}
%
\section{Conclusion}
\label{sec:conclusion}
This paper proposes 3R-INN, a single network releasing a minimum viable quality, low-resolution, grain-free and energy-aware image, from an \ac{hr} grainy image. 3R-INN enables to reduce the overall energy consumption in the video transmission chain by reducing the energy needed for encoding, transmission, decoding and display. Furthermore it does not need to transmit auxiliary information to reconstruct the original grainy content, since all the lost information including details, film grain and brightness was encoded and disentangled in a standard Gaussian distribution, through a latent encoding block conditionned on the \ac{lr} image. As it performs 3 tasks at once, with a single network of less than 2M parameters, 3R-INN also reduces the total processing energy of running 3 separate networks, with higher number of parameters. Experimental results demonstrate that 3R-INN outperforms the existing methods by a large margin for film grain synthesis, and achieves state-of-the-art performance in the rescaling and energy-aware tasks. However, for the latter, a fine-tuning for each value of energy reduction rate target $R$ was conducted. Conditioning the network on $R$ to avoid fine-tuning different networks for each value of $R$, will therefore be investigated in the future. Some subjective test will also be conducted to assess the acceptability by end users of the provided LR energy-aware images.

{\small
\bibliographystyle{ieee_fullname}
\bibliography{egbib}
}

\end{document}